\documentclass[preprint]{aastex} 

\slugcomment{Revised: 30 May 2008}
\slugcomment{Accepted: 14 June 2008}
\shorttitle{2MASS Ultracool Dwarfs}
\shortauthors{Reid et al.}

\begin{document}
\def\pedant{Mart{\'{\i}}n }
\def\etal{{\sl et al.}}
\def\etall{{\sl et al. }}
\def\pma{$\arcsec$~yr$^{-1}$ }
\def\kms{km~s$^{-1}$ }
\def\msun{$M_{\odot}$}
\def\rsun{$R_{\odot}$}
\def\lsun{$L_{\odot}$}
\def\halpha{H$\alpha$}
\def\hbeta{H$\beta$}
\def\hgama{H$\gamma$}
\def\hdelta{H$\delta$}
\def\Teff{T$_{eff}$}
\def\logg{$log_g$}

\title{Meeting the Cool Neighbors X: Ultracool dwarfs from the 2MASS All-Sky Data Release}

\author{I. Neill Reid\altaffilmark{1}}
\affil{Space Telescope Science Institute, 3700 San Martin Drive, 
Baltimore, MD 21218; inr@stsci.edu}

\author{ Kelle L. Cruz\altaffilmark{1, 2}}
\affil{Department of Astronomy, MC 105-24, California Institute of Technology,  kelle@astro.caltech.edu }

\author{J. Davy Kirkpatrick\altaffilmark{1}}
\affil{IPAC, California Institute of Technology, Pasadena, CA 91125}

\author{Peter R. Allen\altaffilmark{1}}
\affil{Pennsylvania State University, Dept.\ of Astronomy and Astrophysics, 525 Davey Lab, University Park, PA 16802; pallen@astro.psu.edu}

\author {F. Mungall\altaffilmark{1}}
\affil{Department of Physics and Astronomy, University of Pennsylvania, 
209 South 33rd Street, Philadelphia, PA  19104}

\author{James Liebert\altaffilmark{1}}
\affil{Department of Astronomy and Steward Observatory, 
University of Arizona, Tucson, AZ 85721}

\author{Patrick Lowrance\altaffilmark{1}}
\affil{IPAC, California Institute of Technology, Pasadena, CA 91125}

\author{ Anne Sweet}
\affil{Department of Astronomy, MC 105-24, California Institute of Technology}

\altaffiltext{1}{Visiting Astronomer, Kitt Peak National Observatory, 
NOAO, which is operated by AURA under cooperative agreement with the NSF.} 
\altaffiltext{2}{Spitzer Fellow}

\begin{abstract}

Using data from the 2MASS All-Sky Point Source Catalogue, we have extended our census of nearby ultracool dwarfs to cover the full celestial sphere above Galactic latitute 15$^o$. Starting with an initial catalogue of 2,139,484 sources, we have winnowed the sample to 467 candidate late-type M or L dwarfs within 20 parsecs of the Sun. Fifty-four of those sources already have spectroscopic observations confirming them as late-type dwarfs. We present optical spectroscopy of 376 of the remaining 413 sources, and identify 44 as ultracool dwarfs with spectroscopic distances less than 20 parsecs. Twenty-five of the 37 sources that lack optical data have near-infrared spectroscopy. Combining the present sample with our previous results and data from the literature, we catalogue 94 L dwarf systems within 20 parsecs. We discuss the distribution of activity, as measured by H$\alpha$ emission, in this volume-limited sample. We have coupled the present ultracool catalogue with data for stars in the northern 8-parsec sample and recent (incomplete) statistics for T dwarfs to provide a snapshot of the current 20-parsec census as a function of spectral type. 

\end{abstract}


\mbox{}

\section{Introduction}

The closing years of the twentieth century saw the completion of the first large-scale, deep, near-infrared sky surveys, DENIS (Epchstein \etal, 1994) and 2MASS (Skrutskie \etal, 2006). Results from those surveys, and from the deep optical imaging of the Sloan Digital Sky Survey (York \etal, 2000), have revolutionised our understanding of the very low-mass dwarfs that populate the lower reaches of the HR diagram. Although predated by the identification of the first incontrovertible brown dwarf (Nakajima \etal, 1995), the avalanche of discoveries over the past decade (Delfosse \etal, 1997; Kirkpatrick \etal, 1999, 2000 - hereinafter, K99, K00; Fan \etal, 2000; Hawley \etal, 2002) would not have been possible without the unparalleled sensitivity provided by those surveys. Initial investigations operated in discovery mode, pushing detection to lower and lower temperatures, and extending the spectral classification system to types L (K99;Mart{\'{\i}}n \etal, 1999) and T (Geballe \etal, 2002; Burgasser \etal, 2003, 2006). At the same time, analyses of ensemble properties of observational samples, combined with detailed studies of individual objects, have resulted in greater insight into their evolution, atmospheric structure and composition (Baraffe \etal, 1998; Burrows \etal, 2001; Marley \etal, 2002). 

Understanding the statistical properties of brown dwarfs requires that we move beyond discovery mode, and define reliable, unbiased catalogues of late-type dwarfs. As part of the NASA/NSF NStars initiative, we have been undertaking a systematic survey for M and L dwarfs within 20 parsecs of the Sun. Our initial efforts centred on the 48\% of the sky covered by the 2MASS Second Incremental Release (the 2MASS IDR2) and the results from those studies are described in previous papers in this series. We have adopted two main strategies to exploit data from the 2MASS IDR2. 

First, we have cross-referenced 2MASS data against the NLTT catalogue of proper motion stars (Luyten, 1980), and used the resulting optical-infrared colours to identify early- and mid-type M dwarfs within 20 parsecs of the Sun. The main results from our M-dwarf surveys are summarised in Papers VIII and XI of this series (Reid et al, 2004; 2007), which present preliminary J-band luminosity functions, $\Phi(M_J)$, for stars within 20-parsecs of the Sun. That dataset includes over 1100 early and mid-type M dwarfs in $\approx$1000 systems within 20 parsecs of the Sun. We are using a variety of techniques to improve the completeness of this sample, including spectroscopic follow-up of additional nearby-star candidates from the recent proper motion surveys undertaken by L\'epine \& Shara (2005) and L\'epine (2008). At least 450 systems in the current census lack trigonometric parallax data, while at least half of the stars have not been scrutinised for spectroscopic or astrometric binary companions. Given the substantial numbers in this sample, obtaining those ancillary data has higher priority than extending the M dwarf survey beyond the bounds of the 2MASS IDR2 database. We refer to these M dwarfs as the 2M2nd sample.

Second, we have used 2MASS photometry to search directly for ultracool dwarfs - spectral types M7 to L8. Paper V (Cruz et al, 2003) and Paper IX (Cruz et al, 2007) summarise the techniques used to define ultracool candidates in the 2MASS IDR2 (the 2MU2 sample), and outline the main results from our analysis of that sample. In total, we identified 637 candidate nearby ultracool dwarfs, and have accumulated optical spectroscopy of 480 of those objects. Three hundred and eighty-nine are confirmed as spectral type M7-L6, including 277 new identifications. A future paper in this series will present analysis of near-infrared spectra of the faintest ultracool dwarfs from the 2MU2 sample. Combining these results gives the first volume-complete sample of L dwarfs, and the first derivation of the luminosity function for spectral types M8 to L8. 

In contrast, the 2MU2 20-parsec sample includes only 89 ultracool dwarfs, comprising 49 late-M dwarfs and 40 L dwarfs. Those sparse statistics, combined with general interest in the intrinsic properties of these cool, very low-mass dwarfs, provide strong incentive to expand our survey. With the release of the 2MASS All-Sky Survey, we have the opportunity to double the areal coverage of our investigation. This paper summarises the results of that process.

We have used the experience gained in compiling the 2MU2 sample to refine the selection criteria, and focus our candidate list with higher efficiency on {\sl bona-fide} ultracool dwarfs. As in our previous analyses, we use optical spectroscopy as the prime tool for verifying the nature of the candidates, and estimating distances to confirmed ultracool dwarfs. The far-red spectra also allow us to identify lower-mass brown dwarfs, {\it via} the presence of lithium absorption, and active objects with appreciable H$\alpha$ emission. 

The present paper is organised as follows: Section 2 describes the revisions made to the selection criteria used to construct the 2MUA sample, and summarises the broad properties of the initial candidate list; Section 3 describes follow-up optical spectroscopy and the spectral classification of candidates; Section 4 discusses the distance distribution and overall properties of the sample, as well as describing some of the more unusual objects in the sample; and Section 5 summarises our conclusions.

\section {The All-Sky Ultracool Sample}

The 2MU2 ultracool sample discussed in Papers V and IX is drawn from the 2MASS IDR2, which covers 48\% of the sky. Those data were refined for inclusion in the 2MASS All-Sky Point Source catalogue, which forms the basis for our present analysis. As a result, there is significant overlap between the ultracool candidates identified here and the previous 2MU2 sample. For clarity, we treat these two datasets separately, and refer to the new candidates as the 2MUA sample. 

\subsection {Defining the 2MUA sample}

The primary criteria used to define the 2MU2 and 2MUA samples are tied to the near-infrared photometric properties and Galactic location. Drawing from the experience gained in compiling follow-up observations of the 2MU2 sample, we have modified these selection criteria in certain important respects:
\begin{itemize}

\item First, we raised the galactic latitude criterion from $|b| > 10^o$ to $|b| > 15^o$.  Regions near the Galactic Plane suffer from two major problems for our type of survey: high source density, leading to incompleteness and photometric inaccuracies due to image crowding; and extensive reddening, due to interstellar dust. To minimise the effects on the 2MU2 sample, we excluded all 2MASS IDR2 tiles that are centred at Galactic latitudes $|b| < 10^o$. This had a relatively small impact on areal coverage, since the 2MASS IDR2 release covered predominantly high galactic latitudes. Nonetheless, the 2MU2 ultracool candidates included a significant number of reddened sources. With the higher latitude limit adopted in the present analysis, the majority of those sources are eliminated {\sl a priori}. The $|b| > 15^o$ requirement reduces coverage to $\sim70\%$ of the celestial sphere.

\item Second, we have increased the blue (J-K$_S$) limit at bright magnitudes from (J-K$_S) > 1.00$ to (J-K$_S) > 1.05$. The 2MU2 candidate list includes several hundred sources with J$>$12 and (J-K$_S) \le 1.05$ (Figure 4 in Paper V), almost all of which have proven to be M6/M6.5 dwarfs at distances of 30-50 parsecs. Eliminating the M6 dwarfs comes at a price: with the redder colour limit, the 2MUA sample includes few M7 and only a subset of nearby M8 dwarfs. Based on the spectral-type/colour distributions derived by Gizis \etall (2001), we expect approximately 50\% of M8 dwarfs to meet the current colour limits.

\item Third, we consider only sources with magnitudes $J > 9$. Five hundred and eighty-eight of the 2MU2 candidates are brighter than this limit, but only four of these sources proved to be mid- or late-type M dwarfs. 

\end{itemize}
Other selection criteria outlined in Paper V, based on location in the JHK$_S$ (including the `giant star' criteria defined in equation (4) of that paper), on location in the J/(R-J) diagrams and on the 2MASS photometric confusion/solar system flags ($cc_{flg}=000$, $mp_{fig}=00$), were retained unaltered.

The upper section of Table 1 (corresponding to Table 1 in Paper V) breaks down the steps used to construct the 2MUA candidate list. As with the 2MU2 sample, the initial catalogue of 2.14 million 2MASS sources with (J-K$_S) > 1.05$ and latitude $|b| > 15^o$ is reduced to manageable proportions, primarily through cuts in the (J, (J-K$_S$)) and (J-H)/(H-K$_S$) planes, and the elimination of sources in highly-crowded fields near the $|b|=15^o$ cutoff or near known star-forming regions. As discussed in Paper V, rough positions and dimensions for highly-reddened regions were taken from Dame et al (1987) and Dutra \& Bica (2002), enlarged as necessary where visual inspection revealed high source densities around the edges of the excised regions. Sources eliminated based on this criterion are designated as `clouds/crowded' in Table 1. 

The 2MASS All Sky Point Source Catalogue includes the 48\% of the sky covered by the 2MASS IDR2. Our prior analyses, described in Papers V and IX, resulted in the identification of 1672 ultracool candidates within those regions (Paper V, Table 1), and follow-up observations have confirmed 369 as ultracool dwarfs. Eliminating the 2MU2 sources from the all-sky sample gives a total of 1,018 candidate ultracool dwarfs in the 2MUA sample.  Figure 1 plots the ($\alpha, \delta$) and ($l, b$) distributions of the 2MU2 and 2MUA candidates. Combined, the two datasets cover approximately 26,500 square degrees or $\sim65\%$ of the sky.

2MASS and Digital Sky Survey images of 1,018 ultracool candidates in the 2MUA sample were inspected individually, and the results from those inspections are given in Table 1. Our inspection showed that almost half the sample proved to be artefacts, mainly diffraction spikes and blends with background nebulosity. (The 2MASS catalogue includes a number of flags to identify sources of dubious image quality, and almost all of these sources are indeed flagged as suspect.) A further 26 sources were eliminated since they lie near small star-forming regions (``clouds" in Table 1), and a similar number were disqualified because they lie within 15$^o$ of the Galactic equator (the phase I cuts are based on the Galactic latitude of the centre of the 2MASS tile, not the positions of the individual objects in each tile). Thirty-one objects have optical/IR colours (based on DSS data) that are obviously inconsistent with ultracool dwarfs, and, finally, eight objects are artefacts associated with bright (HD catalogue) stars. Removing those sources reduces the candidate list to 467 sources.

\subsection {Known ultracool dwarfs}

Over seventy sources in our candidate list have been observed in the course of other surveys for ultracool dwarfs. Fifty-four stars (and brown dwarfs) have extant optical spectroscopy of sufficient signal-to-noise and resolution to allow unambiguous classification as late-type dwarfs; in a few cases, we have supplemented the literature data with our own observations. The relevant characteristics of these objects are given in Table 2. Most were identified from follow-up observations of extremely red sources from the 2MASS survey (K00; Gizis et al, 2000; Gizis et al, 2002), the DENIS survey (Mart{\'{\i}}n et al, 1999; Phan-Bao et al, 2001) and the SDSS survey (Fan et al, 2000; Hawley et al, 2002).  Twenty-one dwarfs listed in Table 2 have trigonometric or spectroscopic parallaxes that indicate distances within 20 parsecs of the Sun. 

\section {Spectroscopy}

\subsection {Observations}

We have obtained intermediate-resolution optical spectroscopy of 376 sources from the 2MUA sample. The overwhelming majority of the observations, covering some 355 sources, were obtained in the course of several observing runs between March 2003 and February 2004. We used the RC spectrograph on the Kitt Keak National Observatory 2.1-metre telescope in March and October 2003; the MARS spectrograph on the KPNO 4-metre telescope in July 2003 and February 2004; the RC spectrograph on the 1.5-metre telescope at Cerro-Tololo Interamerica Observatory in May and November 2003; and the RC spectrograph on the CTIO 4-metre telescope in April 2003, August 2004 and January 2006. In each case, the spectra cover the wavelength range 6300 - 10000\AA\ at a resolution of $\sim7$\AA. 

Twenty-one fainter candidates were observed using the Gemini telescopes. The Gemini Multi-Object Spectrometer (Hook \etal, 2004) was used on Gemini North (GN) and South (GS) during queue observations taken between August 2004 and November 2005 (Program IDs: GN-2004B-Q-10, GS-2004B-Q-30, GN-2005B-Q-20, GS-2005B-Q-21). The observations were made using the RG610\_G0307 filter and R400\_G5305 disperser on Gemini North, while the RG610\_G0331 filter and R400\_G5325 disperser were used on Gemini South. In both cases, the data cover the wavelength range 6000--10000~\AA. Two consecutive observations, with the central wavelength offset, were taken of each target to provide complete wavelength coverage. On both telescopes, the nod and shuffle mode was used with a 0\farcs75-wide slit to provide good sky subtraction and a resolution of 5.5~\AA\ (4 pixels).

The spectroscopic data acquired from the KPNO and CTIO telescopes were bias-subtracted and flat-fielded using the IRAF CCDRED package, and the spectra extracted, wavelength- and flux-calibrated using standard techniques. The wavelength calibration is based on a single HeNeAr arc, usually taken at the start of the night. Each night we also observed one of the following flux standards: BD+26 2606, BD+17 4708, HD 19445 (from Oke \& Gunn, 1983); Feige 56, Feige 110 or Hiltner 600 (from Hamuy et al, 1994). 

The Gemini GMOS package was used to reduce the data from Gemini North and South\footnote{Our reduction methods for both traditional and Nod and Shuffle GMOS data are described in detail at http://www.astro.caltech.edu/$\sim$kelle/gmos/gemini\_NSreduction.html.}. Nod and shuffle dark frames were subtracted and the data were flat fielded using the \textit{gsreduce} task and the sky lines were subtracted using \textit{gnsskysub}. Flux calibration was provided through observations of the flux standards G191B2B, LTT 1020 and EG 21 (Massey \etal, 1988; Massey \& Gronwall, 1990; Hamuy \etal, 1994). All spectra were extracted using \textit{gsextract} and the flux calibration applied with \textit{calibrate}. As discussed in Cruz \etall (2007), the slope of the spectra from GN-2004B-Q-10 are systematically too steep longward of 8700~\AA. None of the spectra have been corrected for telluric absorption.  

The 2MASS sources targeted in these observations are listed in Tables 3 to 6, where the coordinates and near-infrared photometry are from the 2MASS All Sky Point Source Catalogue, and the results deduced from the observations. Table 3 lists data for 44 sources that we identify as ultracool dwarfs likely to lie within 20 parsecs of the Sun; Table 4 presents data for 228 ultracool dwarfs at larger distances; Table 5 lists 83 spectroscopically confirmed K and M giants; and Table 6 catalogues data for 22 carbon stars. Combining literature data and our own observations, we have optical spectra for 430 of the 467 ultracool candidates. Twenty-eight of the remaining 37 sources have been observed spectroscopically at near-infrared wavelengths, and nine sources have no follow-up observations. Twenty-five of the 28 sources with observations have spectra consistent with ultracool dwarfs lying more than 20 parsecs from the Sun. Those infrared observations will be discussed in detail in a future paper in this series.

\subsection {Spectral types and distances}

We have applied the methods described in Papers V and IX to determine spectral types, and hence absolute magnitude and distance estimates, for the dwarfs in the 2MUA sample. As discussed in those papers, although molecular bandstrengths are well correlated with luminosity for early and mid-type M dwarfs, there are ambiguities for later-type dwarfs. We therefore determine spectral types for the latter dwarfs from the overall spectral energy distribution from 6,000 to 10,000 \AA, using side-by-side comparison with spectral standards. The uncertainties are generally $\pm0.5$ subtypes for well-exposed spectra, rising to $\pm1-2$ subtypes for low signal-to-noise data. In cases where we have multiple observations of a particular candidate, we have used the highest S/N spectrum to estimate the spectral type.

Absolute magnitudes of late-type dwarfs ($>$M6) are derived directly from the spectral types using the calibration given in Paper V. For earlier-type dwarfs, we derive distances using the TiO5, CaH2 and CaOH bandstrengths and the relations listed in Paper III. In both cases, the calibrations are tied to the 2MASS J passband. The results are collected in Tables 3 and 4. Table 3 presents observations of 44 ultracool dwarfs (spectral types M7 and later) with formal distances less than 20 parsecs. Table 4 lists data for a further 228 dwarfs that lie beyond our distance limit, including 84 L dwarfs and 132 ultracool M dwarfs. Many late-type M dwarfs and a handful of L dwarfs exhibit H$\alpha$ emission. Schmidt et al (2007) present a thorough analysis of chromospheric activity in these low-mass dwarfs, and also discuss the proper motions and correlations between activity and kinematics. Detailed discussions of individual objects of interest are given in \S4.

As in our previous spectroscopy of 2MU2 ultracool candidates, a number of late-type dwarfs exhibit anomalously strong VO absorption and/or weaker K I and Na I atomic absorption. This is generally interpreted as evidence for surface gravities that are lower than the typical values for field dwarfs of similar spectral types (Kirkpatrick \etal, 2006). Twenty-seven candidate low gravity dwarfs are identified in Tables 2, 3 and 4. We have assigned these objects spectral types and, where necessary, spectroscopic parallaxes using conventional criteria; however, if the systems prove to be young, low gravity dwarfs, it is likely that both spectral types and distances will require revision. Consequently, the spectral types for these objects are listed in parentheses in Tables 2-4. The full characteristics of these candidate low gravity dwarfs will be discussed in more detail by Cruz et al (in preparation).

Finally, late-type first giant branch and asymptotic giant branch stars can have near-infrared colours that meet our selection criteria, and our follow-up spectroscopic observations have identified a number of such stars. Tables 5 and 6 list data for 83 K and M giants and  22 carbon stars, respectively. A number of carbon-rich dwarfs have been discovered through follow-up observations of 2MASS ultracool candidates (e.g. Lowrance \etal, 2003). However, the stars listed in Table 6 all have classical carbon giant spectra, and none shows evidence for significant proper motion. Consequently, it is likely that all are giants rather than nearby dwarfs.

\section {Dwarfs of particular interest}

\subsection {Supplementary spectral standards}

Spectral classification is a comparative technique, where the overall appearance of a program object is matched against a set of reference calibrators. It is therefore important to have well defined standard objects. This is particularly the case for ultracool dwarfs, where spectral type appears to be the empirical parameter that is linked most closely to physical characteristics, such as luminosity and temperature.  

Ideally, spectral standards should be bright objects that are accessible to even moderate-aperture telescopes. The primary L dwarf spectral standards are specified by Kirkpatrick \etall in their definition of spectral class L (Table 6 of K99). At that juncture, only $\sim25$ L dwarfs were known, and, with only a limited parent sample, the later-type standards are relatively faint. Moreover, several of the brightest standards have proven to be close binaries. This is not unexpected, given that this initial set was drawn from a magnitude-limited sample.

Spectroscopic observations have now been obtained for more than 500 L dwarfs, including some that are significantly brighter (in apparent magnitude) than the primary standards in the initial sample. In particular, the present survey, which concentrates on the nearest (and therefore the brightest) L dwarfs, provides an excellent resource for supplementing the reference set of primary standards. All of these observations are catalogued in the on-line L dwarf database maintained at http://DwarfArchives.org. 

We have selected supplemental spectral standards based on three criteria: apparent brightness; the absence of a known close companion; and spectral morphology. We have not given consideration to the declination of the source (i.e. accessibility from northern and southern ground-based observatories). All bright ($J\la14$) objects of each subclass that are currently not known to be binary were considered initially. The candidate standards were matched against the original standards through overplotting the spectra, and by comparing the four spectral indices (CrH-a, Rb-b/TiO-b, Cs-a/Vo-b and color-d) used for spectral typing in K99. Indices for the original standards and new candidates were measured using the same script; our measurements reproduce the values reported in K99 for the original standards. Table 7 catalogues both the original standards and the new objects that best match the original classification scheme, both quantitatively (via spectral indices) and qualitatively (via overplotting). Figure 2 shows how the spectral indices measured for the new standards compare with the primary sequence; and Figure 3 directly compares the far-red optical spectra of the primary and supplementary standards.

We have not been able to identify any completely acceptable L6 or L7 type supplemental standards from the present observational dataset. In order to confidently choose a spectral standard, a spectrum of fairly high signal-to-noise is required. These late-type L dwarfs are of low luminosity and few objects in our library have spectra of sufficient quality to enable a reliable comparison with the original standard. (Note that we are fortunate that the new L8 standard, with a distance (just) less than 5~pc, is one of the closest brown dwarfs known.) However, we identify 2MASS J15150083+4847416 and 2MASS J09083803+5032088 as potential L6 and L7 standards, respectively. Higher signal-to-noise data than our current observations are required before those dwarfs can formally be confirmed as secondary standards

All of the new standards except LSR J0602+3910 have been imaged with NICMOS as part of our search for low-mass companions; none is resolved as a binary (Reid et al, 2006; 2007). We do not have spectrum for the new L4 standard 2MASS~0500+0330 that extends far enough into the red for a color-d index to be measured; however, in all other respects, the object meets the criteria that define a spectral standard. 

\subsection {2M2139+0220: a very early-type T dwarf }

2MASSJ 21392676+0220226 is a faint source (J=15.26) with red near-infrared colours ((J-H)=1.10, (H-K$_S$)=0.58). The prime aim of the present survey is the identification of late-type M dwarfs and L dwarfs, and those colours are broadly consistent with a mid-type L dwarf at a distance of 25-30 parsecs. However, the optical spectrum is smooth and largely featureless, with the exception of absorption by Cs I at 8521 and 8963\AA, and H$_2$O at 9300\AA\ (Figure 4). Moreover, Burgasser \etall (2006) have obtained low resolution, near-infrared spectra that indicate the presence of methane absorption. This shows that 2M2139+0220 is a nearby early-type T dwarf, with spectral type  $\approx$T1.5. On that basis, we estimate a distance of $\sim15$ parsecs. Further near-infrared spectroscopy of this dwarf will be particularly interesting.

\subsection {Lithium and H$\alpha$ detections}

It is now well established that the presence of lithium absorption in ultracool dwarfs indicates that those objects have substellar masses (Rebolo, Mart{\'{\i}}n \& Magazzu, 1992). The critical temperature for lithium burning is $\sim 2 \times 10^6$K, or $\sim10^6$K cooler than the critical temperature for hydrogen burning. Low mass dwarfs are fully convective; thus, the presence of detectable lithium in the photosphere indicates that the core temperature has never reached the critical value for hydrogen fusion. Theoretical models (Chabrier \& Baraffe, 1997) predict that lithium remains undepleted in brown dwarfs with masses below 0.055$M_\odot$, while lithium is subject to partial depletion in dwarfs with masses in the range $0.055 < {M \over M_\odot} < 0.075$, with the rate of depletion scaling with increasing mass. 

We have examined our optical spectra, and identified lithium absorption in eight L dwarfs in the present sample. The measured equivalent widths for those sources are given in Table 8. We also list new observations of a number of L dwarfs from the 2MU2 sample. This represents a very low detection rate for the current sample, which is likely to be explained by the spectral resolution of our observations, coupled with the moderate signal-to-noise of our spectra of many late-type L dwarfs. It is notable that all of the lithium dwarfs listed Table 8 were observed using GMOS on Gemini. Higher resolution and higher S/N data are likely to reveal Li 6708\AA\ in a number of other dwarfs in both the present sample and the 2MU2 sample.

Turning to the data listed in Table 8, in most cases the lithium lines are moderately strong, with equivalent width of 3-4\AA. This may indicate that lithium is partly depleted in those systems, suggesting a mass close to 0.065$M_\odot$. There are, however, a handful of dwarfs with much stronger lithium absorption, such as 2M0310-2756, 2M0652+4710 and 2M2317-4838. We also note that several dwarfs listed in this table have spectral signatures consistent with low surface gravity (the spectral types for those dwarfs are enclosed in parentheses). The presence of lithium clearly adds further weight to the hypothesis that these are young, low mass brown dwarfs.

Our optical spectra also allow us to probe chromospheric activity through measurements of H$\alpha$ emission. The overall statistics for activity among ultracool dwarfs are discussed by Schmidt \etall (2007), and we consider the 20-parsec L dwarfs in \S 5. Here, we draw attention to two particularly active dwarfs in the present sample: 2M0407+1546 and 2M1022+5825. The latter dwarf, which is discussed by Schmidt \etall (2007), is an L1 dwarf that shows substantial (order of magnitude) variations in the H$\alpha$ line strength on a timescale of 1-2 days. The L3.5 2M0407+1546, on the other hand, has only one optical observation, with Gemini North, but that observation shows H$\alpha$ emission with an equivalent width of $\sim$60\AA. This makes 2M0407+1546 one of the latest type dwarfs to show substantial chromospheric activity. Further observations may shed light on why this particular dwarf has maintained such a high level of activity at this juncture in its spectral evolution.

\section {A 20-parsec ultracool census}

The primary goal of the present program is to compile a census of ultracool dwarfs within 20 parsecs of the Sun. The combined 2MU2 and 2MUA samples are drawn from $\approx65\%$ of the celestial sphere, excluding regions within 15 degrees of the Galactic Plane and high confusion regions, such as the Magellanic Clouds. As outlined in \S3.2, the follow-up observations described in this paper are not complete: we lack optical observations of 27 of the faintest ultracool candidates. Those sources are most likely to contribute to the lowest luminosity bins in the 20-parsec census. The current results are presented with that caveat in mind, and we defer full analysis of $\Phi (M_J$) to a later paper. 

Combining the 2MUA and 2MU2 datasets gives a total of 196 ultracool dwarfs (M7 to T2.5) with formal distances within 20 parsecs of the Sun. Figure 5 shows the distribution of (J-K$_S$) colours as a function of spectral type and the near-infrared (J-H)/(H-K$_S$) two-colour diagram for dwarfs with reliable photometry (that is, excluding known close binary systems). For reference, we include data for the T0 dwarf, 2M2139+0220 (\S4.2).

Focusing on spectral type L, the current 20-parsec census includes 76 systems from the 2MU2 and 2MUA samples. Astrometry and photometry of those systems are listed in Table 9, together with data for an additional 18 systems culled from the literature\footnote{We include wide, easily-resolved companions of earlier-type main-sequence stars, such as Gl 584C and LHS 102Bab, but not close companions, like LHS 2397aB.}. Most of the additions have been identified from follow-up observations of ultracool candidates from the DENIS survey (e.g. Scholz et al, 2002; Phan-Bao \etal, 2008). For consistency, we have used 2MASS photometry and our spectral-type/M$_J$ relation to estimate distances for the latter objects. This can lead to discrepancies between the distances listed in Table 9 and those given in the original discovery paper; for example, the spectroscopic parallax relation adopted by Phan-Bao \etall (2008) leads to distances that are nearer by $\sim5\%$ at L0, $\sim10\%$ at L2 and $\sim5\%$ at L5. 

Table 9 lists data for 107 dwarfs in 94 systems. Two systems require particular comment. 
\begin{description}
\item[2M0805+4812] was originally classified as an L4 dwarf by Hawley \etal (2002) based on optical spectroscopy. However, Knapp \etal (2004) derived a near-infrared spectral type of L9.5. Burgasser (2007b) has shown that these inconsistencies can be resolved if the system is an unresolved binary, comprising an $\sim$L4.5 primary and a $\sim$T5 secondary.
\item[2M1126-5003] was identified by Folkes \etall (2007) in the course of their search for ultracool dwarfs at low galactic latitude ($|b| \le 15^o$). Based on the $\sim1.0 - 1.6 \mu$m spectrum, Folkes \etall (2007) assigned it a near-infrared spectral type of L9$\pm1$\footnote{We note that there is no type L9 in the optical spectral classication system.} and estimated a distance of only 7.2 parsecs. However, subsequent observations by  Phan-Bao \etall (2008) and Burgasser \etall (2008) have shown that the optical spectrum is consistent with an L4/L5 dwarf, albeit with enhanced FeH absorption (at 9896\AA). Burgasser \etall comment that the near-infrared spectrum is unusually blue, which they attribute to the presence of clouds of condensates in the L dwarf atmosphere. We have adopted the spectral type and distance estimate given in the latter paper. 
\end{description}

Figure 6 plots the ($\alpha, \delta$) and ($l, b$) distributions for the 94 L dwarf systems catalogued in Table 9. Fifteen of the eighteen systems drawn from the literature lie at low galactic latitude, outwith the limits of our ultracool dwarf survey. The remaining three systems, the L/T binary, 2M0805 (Burgasser, 2007b)and  the two L8 dwarfs 2M1523 (Gl 384C, Kirkpatrick et al, 2000) and 2M1632 (Kirkpatrick et al, 1999), fall within the area covered by the 2MU2 and 2MUA datasets. However, all three dwarfs have (J-K$_S$) colours that lie blueward of the (J, (J-K$_S$)) selection criteria. As discussed in Paper V, those criteria were chosen to balance sample completeness against a manageable candidate list. Given the overall statistics and the areal coverage of the 2MU2+2MUA samples, it is likely that 25 to 35 L dwarfs within 20 parsecs of the Sun remain to be discovered in the $|b| < 15^o$ Galactic equatorial zone.

Seventy-two of the L dwarf systems within 20 parsecs have been observed using high-resolution imaging techniques. Eleven are resolved as close binary systems, corresponding to a binary fraction of $15.3_{-3.3}^{+5.1}$\% (Reid et al, 2008). As discussed extensively elsewhere (e.g. Burgasser \etal, 2007), almost all ultracool binaries have near-equal mass ratios, and few lie at separations exceeding 15 AU. Data for the companions to the 20-parsec L dwarfs (including five T dwarfs) are given in Table 9.

Figure 7 shows the likely spectral type distribution of dwarfs in the Solar Neighbourhood. The upper panel plots data for the ultracool dwarfs in the current 20-parsec census\footnote{Although spectral types are often quoted at a resolution of 0.5 classes, we have binned the data in unit spectral types since integer types are favoured over half-integral types in our classification process (see Paper V, \S4.1): for example, there are 8 sources classed as L3, but only 3 as L3.5; 6 are classed as L6, but only 2 as L6.5; and 24 are classed as M9, but only 4 as M9.5.}. This is effectively a luminosity function, since we derive absolute magnitudes using the following relation (from Paper V)
\begin{equation}
M_J \quad = \quad -4.410 + 5.04(ST) - 0.6193(ST)^2 + 0.03453(ST^3) - 6.892 \times 10^{-4} (ST^4)
\end{equation}
where ST = 0 for spectral type L0 ($\Delta$ST $\approx 0.55 \Delta$M$_J$). We have identified separately the contribution from known secondary companions. As discussed in Papers V and IX (and \S2.1 of this paper), the initial (J, (J-K$_S$)) colour-magnitude selection criteria lead to the 2MASS ultracool sample becoming incomplete for spectral types earlier than M8 and later than $\sim$L7.

We have extended the spectral type census to the T dwarf r\'egime using the on-line T dwarf database, http://DwarfArchives.org, which currently lists data for 122 T dwarfs. Most lack trigonometric parallaxes, so we have used the (M$_K$, spectral type) relation derived by Burgasser (2007a) to estimate spectroscopic parallaxes. This dataset is highly incomplete, even more so that the late-type L dwarfs, and particularly for the neutral-coloured, early-type T dwarfs\footnote{We note that examples of spectral type T3 are particularly sparse, with only seven dwarfs classified as T3 or T3.5 in entire DwarfArchives database. This compares with 11 T0s and 13 T1s. The nearest T3 dwarf is 2M1206+2813 at a distance of $\sim19$ parsecs.}. Nonetheless the data provide a guide to the current status in the field. Figure 7 clearly suggests that, after a broad minimum spanning $\sim$L5 to $\sim$T2, there is a rise in number density for later-type T dwarfs. This is in accord with expectation, since theoretical models predict that the rate of cooling of brown dwarfs slows with decreasing temperature, leading to a pile-up in numbers at later spectral types (Allen \etal, 2005; Burgasser, 2005).

The lower panel in Figure 7 provides a broader context by expanding the L/T sample to include the expected contribution of K and M dwarfs to the 20-parsec census. We have estimated the likely numbers of earlier-type dwarfs using the statistics for the northern 8-parsec sample (Reid \etal, 2006), adjusting the numbers to an all-sky 20-parsec survey. We have also scaled the observed numbers of L and T dwarfs by a factor of 1.5 to allow for as-yet undiscovered ultracool dwarfs at low galactic latitudes. The resultant distribution illustrates the dominant contribution made by M dwarfs to the visible stellar populations in the Galactic disk. The expectation is that deep surveys at near- and mid-infrared wavelengths will reveal increasing numbers of cool late-type T dwarfs and even cooler Y dwarfs.

Finally, we note that eighty-seven of the L dwarf systems listed in Table 9 have optical spectra\footnote {The three systems that currently lack such data are 2M0155+0950, 2M0830+4828 and 2M1550-442}. Ten systems (12.5\%) have detectable H$\alpha$ emission. The frequency is clearly higher at earlier spectral types, with eight of the active systems having spectral types in the range L0 to L2, including 6 of the 24 L0/L1 systems (25\%). The latest type dwarf that shows evidence of chromospheric activity is 2M0318-3421, an L7 dwarf at a distance of $\sim16.5$ parsecs. 

\section {Summary and conclusions}

As part of our continuing survey of the ultracool dwarfs in the immediate
Solar Neighbourhood, we have used the 2MASS All-Sky Database to extend coverage to all regions of the sky with galactic latitudes $|b| > 15^o$. We have identified 467 candidate nearby ultracool dwarfs, and this paper presents literature data and our own optical spectroscopic observations of 430 of those candidates. Of this subset, sixty-five dwarfs have formal distances within 20 parsecs of the Sun, including 44 observed here for the first time. Examining the full dataset, we have identified several dwarfs with lithium absorption, indicating masses less than $\sim$0.065$M_\odot$

We have combined the present dataset with our previous surveys of K, M and L dwarfs, from Papers V, VIII and IX in this series, and with current census information on nearby T dwarfs from the on-line database, http://DwarfArchives.org, to provide an estimate of the spectral type distribution of late-type dwarfs within 20 parsecs of the Sun. The results show how M dwarfs dominate the local population. The ultracool sample is known to be incomplete for late-L and T dwarfs; nonetheless, the current data show a pronounced minimum from $\sim$L5 to $\sim$T2, with an upturn in the number densities of mid- and late-type T dwarfs. A future paper will present near-infrared spectroscopy of the later-type dwarfs from the present compilation, together with additional sources from the 2MASS All-Sky sample. At that juncture, we will undertake a more quantitative analysis of the ultracool dwarf luminosity function and considering the implications for the mass function in the substellar r\'egime. 

\acknowledgements 
The NStars research described in this paper was supported partially by a grant awarded as part of the NASA Space Interferometry Mission Science Program, administered by the Jet Propulsion Laboratory, Pasadena. 
Support for KLC is provided by NASA through the Spitzer Space Telescope Fellowship Program, through a contract issued by the Jet Propulsion Laboratory, California Institute of Technology under a contract with NASA. PRA. acknowledges support from grant NAG5-11627 to Kevin Luhman from the NASA Long-Term Space Astrophysics program. 
This publication makes use of data products from the Two Micron All Sky Survey, which is a joint project of the University of Massachusetts and the Infrared Processing and Analysis Center/California Institute of Technology, funded by the National Aerospace and Space Administration and the National Science Foundation. We acknowledge use of the NASA/IPAC Infrared Source Archive (IRSA), which is operated by the Jet Propulsion Laboratory, California Institute of Technology, 
under contract with the  National Aerospeace and Space Administration. We also acknowledge making extensive use of the SIMBAD database, maintained by Strasbourg Observatory, and of the ADS bibliographic service. 
This research has made extensive use of the M, L and T dwarf compendium housed at DwarfArchives.org and maintained by Chris Gelino, Davy Kirkpatrick and Adam Burgasser. This program has also profited from extensive allocations of telescope time at both Kitt Peak Observatory and Cerro-Tololo Interamerican Observatory. We thank the NOAO Telescope Allocation Committees for their support of this project and acknowledge the courteous and efficient assistance of the technical support staff: John Glaspey, Darryl Willmarth, Diane Harmer, Bill Gillespie, Hillary Mathis and Hal Halbedel at KPNO; Alberto Alvarez, Angel Guerra and Patricio Ugarte at CTIO.

\clearpage
\centerline {Figure Captions}

\figcaption{ The ($\alpha, \delta$) and ($l, b$) distributions of ultracool candidates from the 2MU2 (red) and 2MUA (cyan) samples. There are 1672 sources in the former sample, and 1018 in the latter (see Table 1).}

\figcaption{Spectral ratios as a function of spectral type for the supplemental
standards listed in Table~7 (\textit{stars}) and the original L
dwarf standards listed in K99 (\textit{circles}).}

\figcaption{L dwarf spectral sequence with supplemental standards (\textit{black})
overplotted on the original standards (\textit{red}). The new data have not been
corrected for telluric absorption, which can significantly affect the spectrum within the shaded region at far-red wavelengths.}

\figcaption{ The 2MASS 2139+0220 (2MUCD 20912 in Table 2), a T0 dwarf at an estimated distance of $\sim14.5$ parsecs.}

\figcaption{ 2MASS near-infrared photometry for the L dwarfs in the 20-parsec sample. We include data for nearby late-M and early-T dwarfs to provide context. The lower panel plots the distribution of (J-K$_S$) colour as a function of the spectral type, including data for the T0 dwarf, 2M2139+0220 (\S4.2).The solid points plot data for early-type T dwarfs from the on-line T dwarf database, http://DwarfArchives.org. The upper panel plots the (J-H)/(H-K$_S$) two-colour diagram for the same dataset, where the symbols match the coding in the spectral type diagram (lower panel). The reddest system is 2M0355+1133, with (J-K$_S$)=2.52 magnitudes. Known binaries are excluded from these diagrams.}

\figcaption{ The ($\alpha, \delta$) and ($l, b$) distributions of the 20-parsec L dwarf sample listed in Table 9. L dwarfs from the 2MU2 (red) and 2MUA (cyan) samples are plotted as crosses; L dwarfs from other datasets are plotted as (green) solid points.}

\figcaption{ The spectral type distribution of stars and brown dwarfs catalogued in the local census. The upper panel plots the distribution of ultracool M and L dwarfs from our 2MASS 20-parsec sample; as discussed in the text, this sample is known to become incomplete for spectral types earlier than M8 and later than $\sim$L7. We also show the spectral type distribution of T dwarfs with distances $d < 15$ parsecs from the on-line T dwarf database, http://DwarfArchives.org, scaling the numbers by a factor of 2 to allow for formal difference the relative volumes sampled (dotted histogram). The T dwarf sample is known to be incomplete. In the lower panel, we combine the ultracool distributions with the spectral type distribution of K and M dwarfs in the northern 8-parsec sample (Reid \etal, 2006; Reid, Cruz \& Allen, 2007; dashed histogram). In each case, the hatched histogram shows the contribution from companions in multiple systems.}

\clearpage

\begin{figure}
\figurenum{1}
\plotone{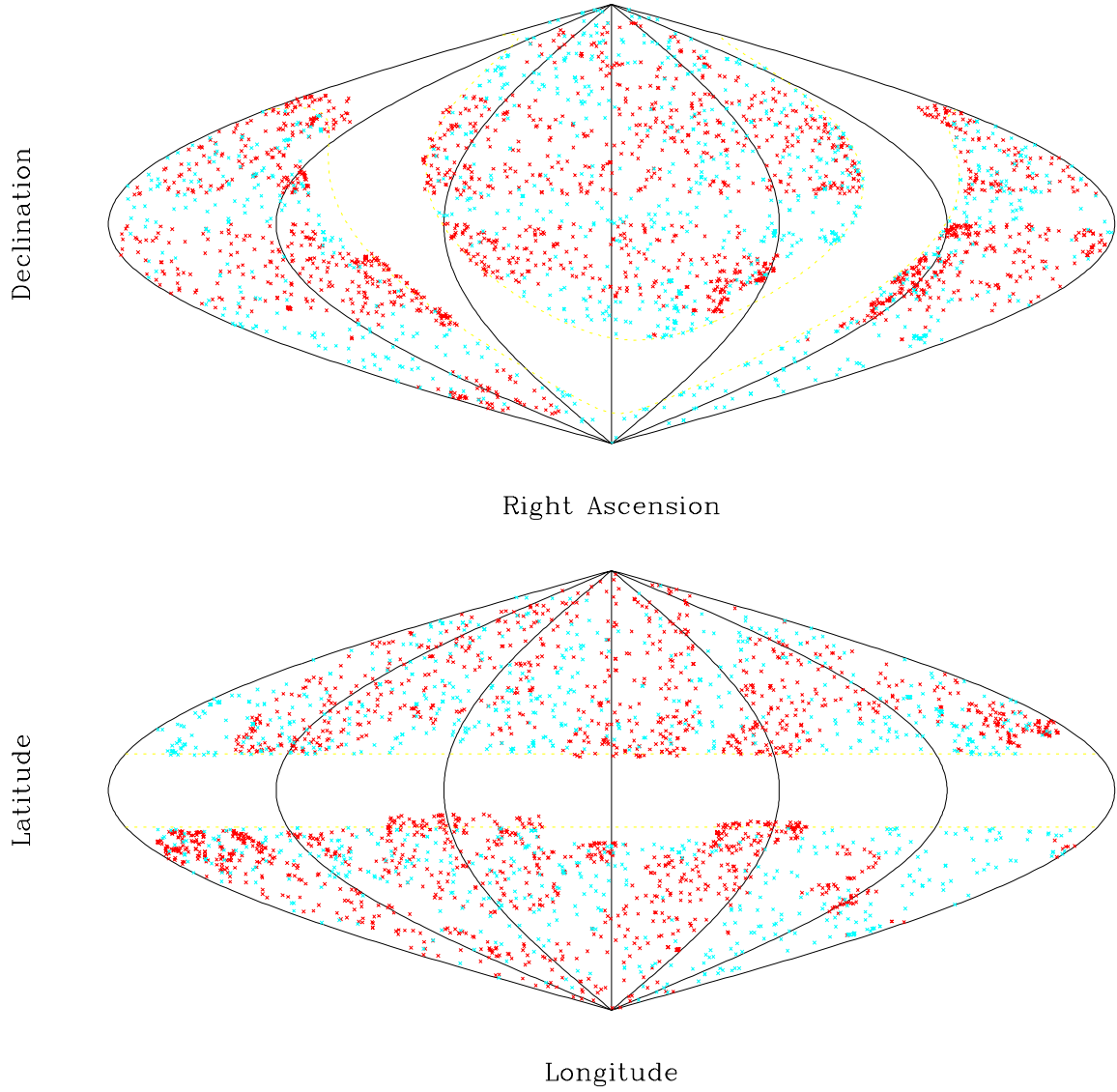}
\caption{ The ($\alpha, \delta$) and ($l, b$) distributions of ultracool candidates from the 2MU2 (red) and 2MUA (cyan) samples. There are 1672 sources in the former sample, and 1018 in the latter (see Table 1).}
\end{figure}
\clearpage

\begin{figure}
\figurenum{2}
\plotone{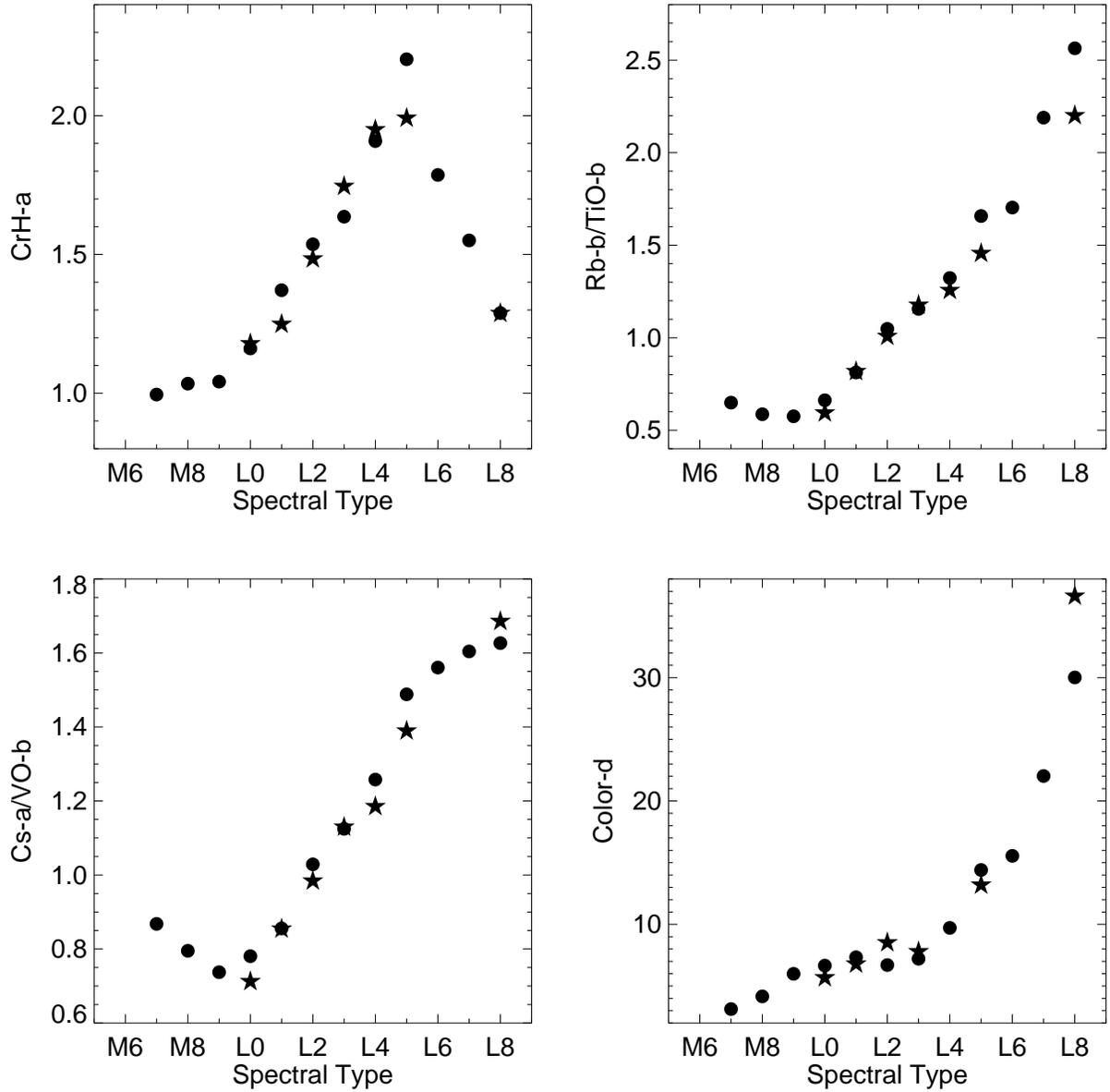}
\caption{Spectral ratios as a function of spectral type for the supplemental
standards listed in Table~7 (\textit{stars}) and the original L
dwarf standards listed in K99 (\textit{circles}).}
\label{fig:indices}
\end{figure}

\begin{figure}
\figurenum{3}
\plotone{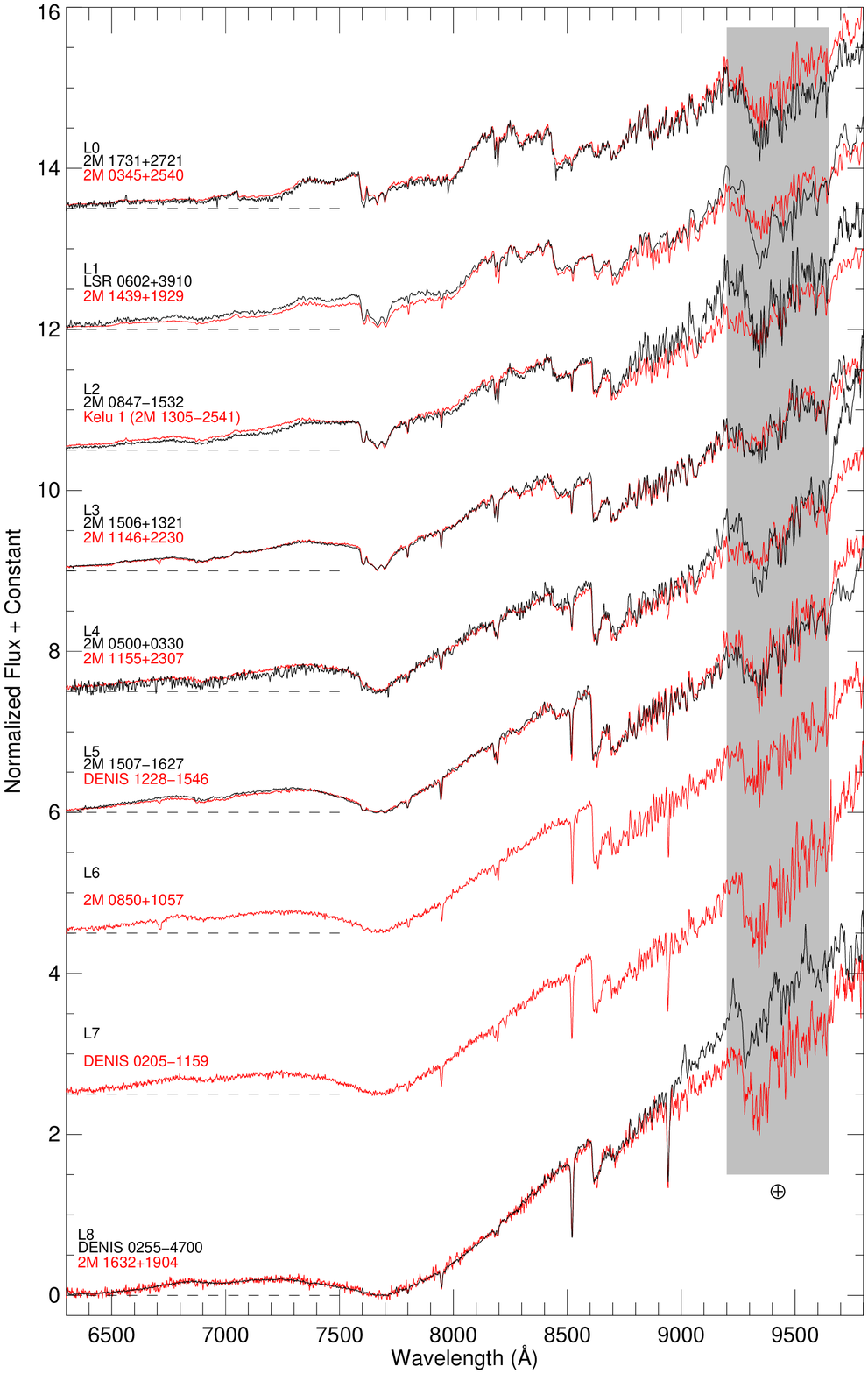}
\caption{L dwarf spectral sequence with supplemental standards (\textit{black})
overplotted on the original standards (\textit{red}). The new data have not been
corrected for telluric absorption, which can significantly affect the spectrum within the shaded region at far-red wavelengths.}
\label{fig:newstds}
\end{figure}

\begin{figure}
\figurenum{4}
\plotone{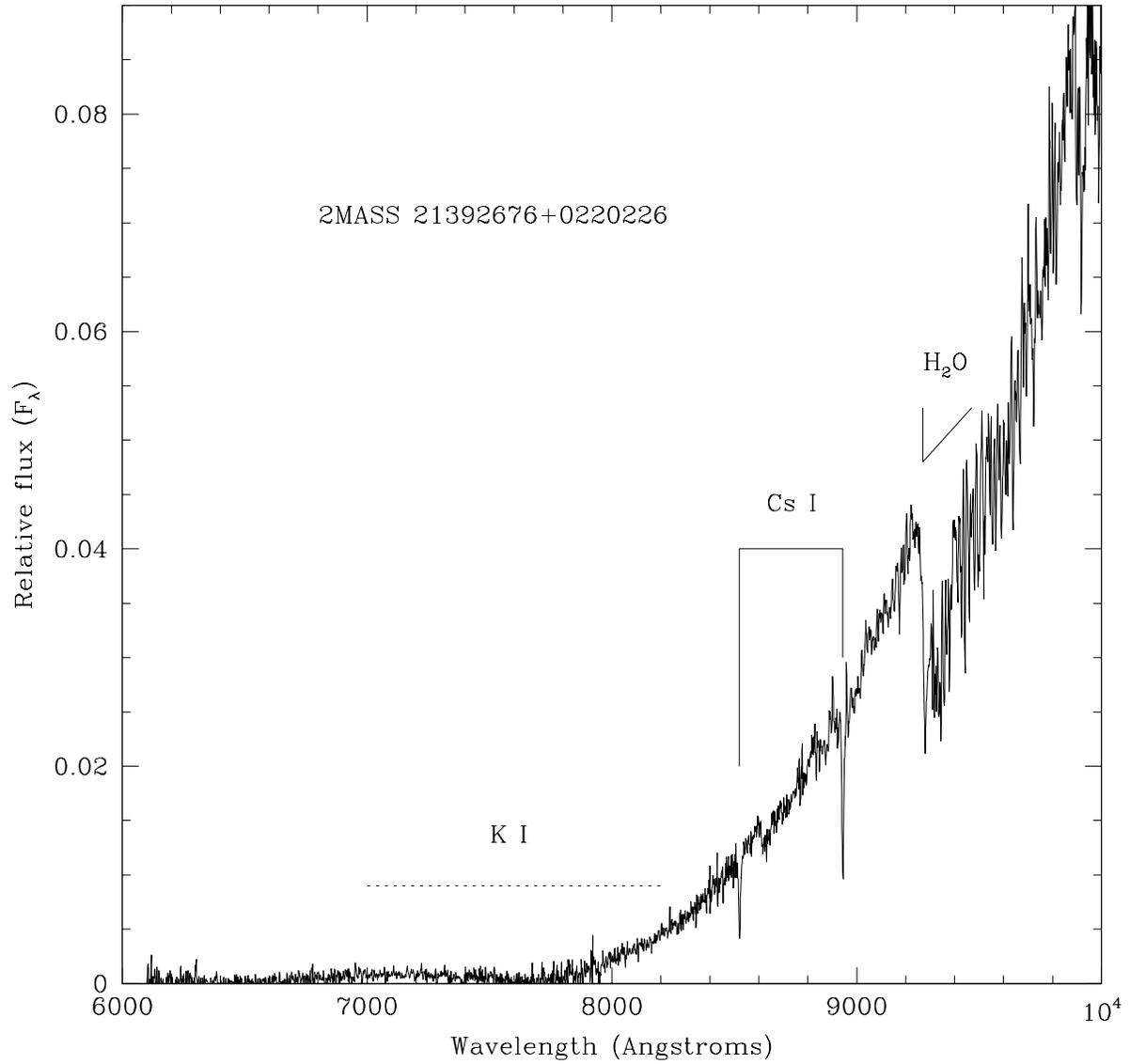}
\caption{ The 2MASS 2139+0220 (2MUCD 20912 in Table 2), a T0 dwarf at an estimated distance of $\sim14.5$ parsecs.}
\end{figure}
\clearpage

\begin{figure}
\figurenum{5}
\plotone{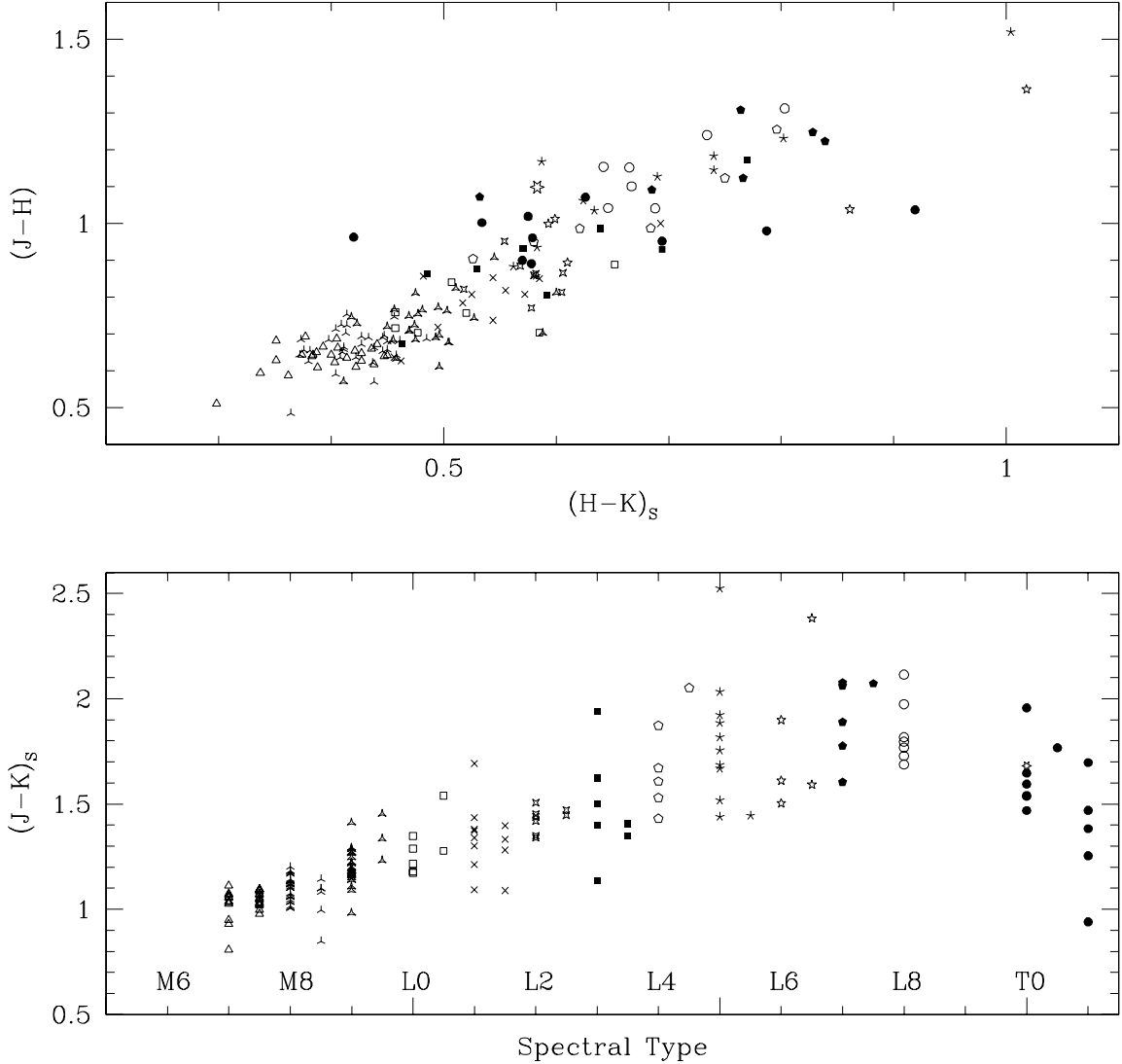}
\caption{ 2MASS near-infrared photometry for the L dwarfs in the 20-parsec sample. We include data for nearby late-M and early-T dwarfs to provide context. The lower panel plots the distribution of (J-K$_S$) colour as a function of the spectral type, including data for the T0 dwarf, 2M2139+0220 (\S4.2). The solid points plot data for early-type T dwarfs from the on-line T dwarf database, http://DwarfArchives.org. The upper panel plots the (J-H)/(H-K$_S$) two-colour diagram for the same dataset, where the symbols match the coding in the spectral type diagram (lower panel).The reddest system is 2M0355+1133, with (J-K$_S$)=2.52 magnitudes. Known binaries are excluded from these diagrams.}
\end{figure}
\clearpage

\begin{figure}
\figurenum{6}
\plotone{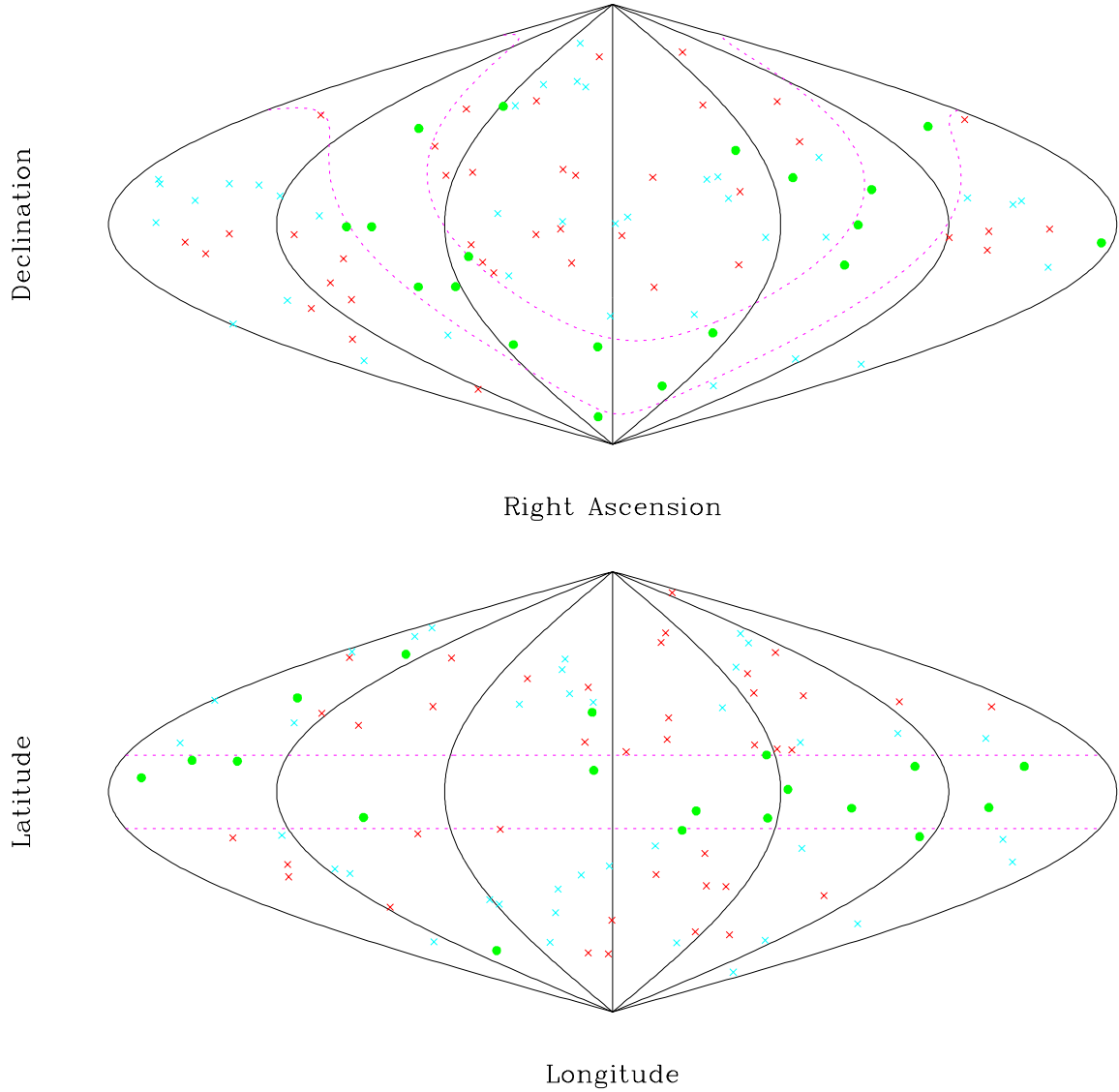}
\caption{ The ($\alpha, \delta$) and ($l, b$) distributions of the 20-parsec L dwarf sample listed in Table 9. L dwarfs from the 2MU2 (red) and 2MUA (cyan) samples are plotted as crosses; L dwarfs from other datasets are plotted as (green) solid points.}
\end{figure}
\clearpage

\begin{figure}
\figurenum{7}
\plotone{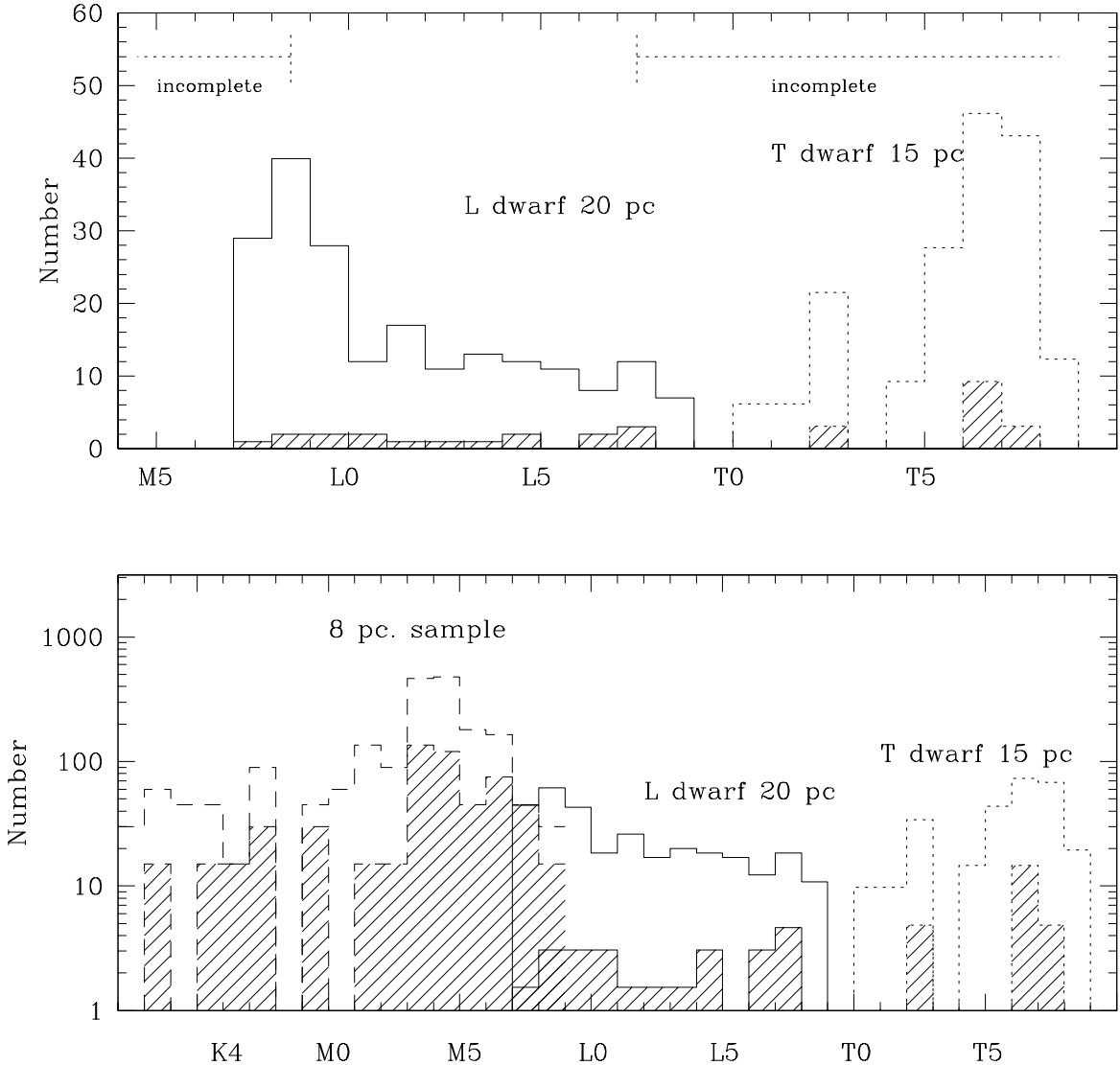}
\caption{ The spectral type distribution of stars and brown dwarfs catalogued in the local census. The upper panel plots the distribution of ultracool M and L dwarfs from our 2MASS 20-parsec sample; as discussed in the text, this sample is known to become incomplete for spectral types earlier than M8 and later than $\sim$L7. We also show the spectral type distribution of T dwarfs with distances $d < 15$ parsecs from the on-line T dwarf database, http://DwarfArchives.org, scaling the numbers by a factor of 2 to allow for formal difference the relative volumes sampled (dotted histogram). The T dwarf sample is known to be incomplete. In the lower panel, we combine the ultracool distributions with the spectral type distribution of K and M dwarfs in the northern 8-parsec sample (Reid \etal, 2006; Reid, Cruz \& Allen, 2007; dashed histogram). In each case, the hatched histogram shows the contribution from companions in multiple systems.}
\end{figure}

\clearpage
\begin{deluxetable}{lr}
\tablewidth{0pt}\tablecaption{Steps to create the 2MUA sample\label{tab:cuts}} \tablehead{\colhead{Item} &
\colhead{Number} } \startdata
\bf{2MASS Hits} & \bf{2,139,482}\\
\hline
Automated cuts \\
\hline
Clouds/crowded & 218,204\\
LMC, SMC, \& 47 Tuc & 76,617\\
M31 \& M33 & 945\\
$J$, ($J-K$) & 1,830,288\\
$(J-H)$, ($H-K$) & 10,462\\
$R$, ($R-J$) & 1,557\\
Giants& 22\\
\hline \hline
All-Sky Total& 1,387\\
In 2MU2 & 369\\
\hline 
\bf{Interim total}& \bf{1,018}\\
\hline
source-by-source cuts \\
\hline
Artefacts & 460 \\
near clouds & 26 \\
$|b| < -15^o$ & 26 \\
blue optical/IR colours & 31 \\
near bright stars & 8 \\
\hline \hline
{\bf Total candidates} & {\bf467} \\
\enddata
\tablecomments{The initial sample of 2.14 million candidates was selected using two criteria: (J-K$_S) > 1.05$ and $|b| > 15^o$. Subsequently, as described in the text, a series of automated cuts were applied to give the All-Sky sample of ultracool candidates. The upper section of this table lists the number of sources rejected based on successive specific criteria (see paper V for full details), reducing the All-Sky sample to to 1387 candidates. Three hundred and sixty-nine of those candidates are included in the 2MU2 sample, so the 2MUA sample consists of 1018 ultracool dwarf candidates. Those candidates were checked on an individual basis (see Section 2.1) and the final list reduced to 467 viable ultracool-dwarf candidates.}
\end{deluxetable}

\begin{deluxetable}{lllrrrcccc}
\tablewidth{0pt}
\tablenum{2}
\tabletypesize{\scriptsize}
\rotate
\tablecaption{Previously Known Cool Dwarfs Recovered in the 2MUF Sample\label{tab:known}}
\tablehead{ \colhead{} & \colhead{} & \colhead{} & \multicolumn{3}{c}{2MASS} & \colhead{}& \colhead{}& \colhead{}& \colhead{}\\
\cline{4-6} & \colhead{} & \colhead{} & \colhead{} & \colhead{} & \colhead{} & \colhead{Optical} \\
\colhead{2MASS Designation\tablenotemark{a}} &  \colhead{2MUCD} & \colhead{Other Names} & \colhead{$J$} & \colhead{$J-H$} &
\colhead{$J-K_S$} & \colhead{Spectral Type\tablenotemark{b}}& \colhead{$M_J$}& \colhead{$d$ (pc)} & \colhead{Refs.\tablenotemark{c}}
}
\startdata
00043484$-$4044058	&	20004 & 	LHS 102B						& 13.109 & 1.054 & 1.713 & L5 \& L5 				& 	\nodata		  					& 11.5	$\pm$ 2.5					&   1, 2\\	%Goldman99*. Golimowski04_SN9*	
00154476+3516026	&	20012 & 		\nodata						& 13.878 & 0.986 & 1.614 & L2						& 12.31$\pm$0.17 					& 20.6	$\pm$ 1.6					&   3, 4\\	%K00*	dahn02* no pi
00210589$-$4244433	&	20018 & 	DENIS-P J0021.0-4244/LEHPM 494B & 13.521 & 0.712 & 1.217 & M9.5						& 11.60$\pm$0.13 					& 24.2	$\pm$ 1.5					&   5, 6\\	%Tinney98*, Basri00*	
00361617+1821104	&	20029 & 		\nodata						& 12.466 & 0.878 & 1.408 & L3.5						& 12.75$\pm$0.03\tablenotemark{d}	& 8.76	$\pm$ 0.06\tablenotemark{d}	&   3, 4\\	%K00-	dahn02-
01075242+0041563	&	20052 & 	SDSS J010752.33+004156.1		& 15.824 & 1.312 & 2.115 & L8						& 14.86$\pm$0.16\tablenotemark{d} 	& 15.59 $\pm$ 1.10\tablenotemark{d}	&   7--9\\	%Schneider02*, Hawley02* Vrba04*
01092170+2949255	&	20055 & 	\nodata							& 12.912 & 0.754 & 1.231 & M9.5						& 11.60$\pm$0.13 					& 18.23 $\pm$ 1.1					&   10 \\	%NN*	
01300580+1721434	&	20070 & 	\nodata							& 13.701 & 0.713 & 1.125 & M8						& 11.16$\pm$0.18 					& 32.3	$\pm$ 2.7					&   10 \\	%NN-	
01353586+1205216	&	20073 & 	\nodata							& 14.412 & 0.885 & 1.494 & L1.5						& 12.15$\pm$0.15 					& 28.4	$\pm$ 2.1					&   3 \\
\enddata
\end{deluxetable}

\begin{deluxetable}{lllllllllllc}
\tablewidth{0pt}
\tablenum{3}
%\tabletypesize{\small}
%\tabletypesize{\scriptsize}
\tabletypesize{\tiny}
\rotate
\tablecaption{M6--L8 Dwarfs Discovered Within 20 pc\label{tab:near}}

\tablehead{ \colhead{} & \colhead{} & \colhead{} &  \multicolumn{3}{c}{2MASS} & \colhead{} & \colhead{} &
\colhead{}& \colhead{} & \colhead{} & \colhead{} \\
\cline{4-6} & \colhead{} & \colhead{} & \colhead{} & \colhead{} & \colhead{} & \colhead{} & \colhead{} & \colhead{Optical}\\
\colhead{2MASS Designation\tablenotemark{a}} &  \colhead{2MUCD} & \colhead{Other Name} &
\colhead{$J$} & \colhead{$J-H$} & \colhead{$J-K_S$} & \colhead{Obs. Date (UT)} & \colhead{Telescope}
& \colhead{Sp. Type\tablenotemark{b}}& \colhead{$M_J$}& \colhead{$d$ (pc)}& \colhead{Refs.\tablenotemark{c}}
}
\startdata
00413538$-$5621127 	& 20035 & DENIS-P J004135.3-562112  & 11.964 	& 0.642 & 1.100 & 2003 Nov 10				& CT 1.5m 	& M8      	& 11.16 $\pm$ 0.18 	& 14.5 $\pm$ 1.2  					&    1, 2 \\% NN4*, PhanBao06*                           \\   
00452143+1634446	& 20037	& \nodata	            	& 13.059	& 1.000	& 1.693	& 2003 Jul 10				& KP 4m		& (L2)		& \nodata			& $\sim$14 							&	3, 22 \\%Wilson03\\
01025100$-$3737438 	& 20049 & LHS 132               	& 11.130 	& 0.651 & 1.061 & 2003 Nov 9				& CT 1.5m 	& M8      	& 11.16 $\pm$ 0.18 	& 12.20 $\pm$ 0.41\tablenotemark{d} &    4, 5, 6 \\%RG05*, Reyle06*. TRIG D Costa05*.  \\   
01090150$-$5100494 	& 20053 & SSSPM J0109-5101      	& 12.228 	& 0.690 & 1.136 & 2003 Nov 9				& CT 1.5m 	& M8      	& 11.16 $\pm$ 0.18 	& 16.4 $\pm$ 1.3					&    7, 8 \\% Lodieu02*, Lodieu05*                       \\   
				 	&  		& \nodata	       			& 		  	& 	  	& 	  	& 2004 Aug 9				& CT 4m   	& 	     	& 				 	& 									&    			                       \\   
01282664$-$5545343  & 20068 & \nodata                   & 13.775 	& 0.859 & 1.439 & 2006 Jan 15				& CT 4m   	& L2      	& 12.31 $\pm$ 0.17 	& 19.6 $\pm$ 1.5					&    9 \\%Kendall07*                                \\   
02150802$-$3040011 	& 20101 & LHS 1367                  & 11.617 	& 0.664 & 1.075 & 2003 Nov 8				& CT 1.5m 	& M8      	& 11.16 $\pm$ 0.18 	& 12.4 $\pm$ 1.0					&    5, 10 \\%Reyle04*,Reyle06-                          \\   
02284243+1639329 	& 20116 & \nodata                   & 13.166 	& 0.840 & 1.348 & 2003 Jul 9,  2004 Feb 11	& KP 4m   	& L0:     	& 11.73 $\pm$ 0.26 	& 19.4 $\pm$ 2.3					&    3 \\
\enddata

\tablenotetext{a}{The sexagesimal right ascension and declination suffix of the full 2MASS All-Sky Data Release designation 
(2MASS Jhhmmss[.]$\pm$ssddmmss[.]s) is listed for each object.
The coordinates are given for the J2000.0 equinox.}

\tablenotetext{b}{Uncertainties on spectral types are $\pm$ 0.5 subtypes except where noted by one or two colons, 
indicating an uncertainty of $\pm$1 and $\pm$2 types respectively. 
Spectra displaying low-gravity features are indicated with parantheses.}

\tablenotetext{c}{References listed pertain to spectral data, resolved binaries, and trigonometric parallaxes.} 

\tablenotetext{d}{Distance estimate based on trigonometric parallax.}
\tablenotetext{e}{Displays variable H$\alpha$ emission as discussed by Schmidt \etal, 2007.}

\tablenotetext{f}{M7: and M8: are estimated to be the component spectral types based on our measured
combined spectral type of M7.5 and $\Delta I \sim1$ found by \citet{Law06}.}

\tablenotetext{g}{$M_J$ estimated using 2MASS $(J-K_S)$ color and the spectral type/$M_K$ relation in \citet{Tinney03}.}

\tablerefs{
(1)~Phan-Bao \etal, 2001;
(2)~Phan-Bao \etal, 2006;
(3)~Wilson \etal, 2003;
(4)~Reid \& Gizis, 2005;
(5)~R\'eyle \etal, 2006;
(6)~Costa \etal, 2005;
(7)~Lodieu \etal, 2002;
(8)~Lodieu \etal, 2005;
(9)~Kendall \etal, 2007;
(10)~R\'eyle \etal, 2004;
(11)~Kirkpatrick \etal, 2008; 
(12)~Henry \etal, 2004;
(13)~Costa \etal, 2006;
(14)~Reid \etal, 2006b;
(15)~Liu, priv. comm..; 
(16)~Bartlettg \etal, 2007;
(17)~Schmidt \etal, 2007;
(18)~Kendall \etal, 2004;
(19)~Law \etal, 2006; 
(20)~Reid \etal, 2006a;
(21)~Reid \etal, 2008;
(22)~Cruz, Kirkpatrick \& Burgasser, in prep.}

\end{deluxetable}

\begin{deluxetable}{lllllllllllc}
\tablewidth{0pt}
\tablenum{4}
\tabletypesize{\scriptsize}
\setlength{\tabcolsep}{0.04in}
\rotate
\tablecaption{Ultracool dwarfs at distances exceeding 20 pc\label{tab:far}}

\tablehead{ 
\colhead{} & \colhead{} & \colhead{} &  \multicolumn{3}{c}{2MASS} & \colhead{} & \colhead{} &
\colhead{}& \colhead{} & \colhead{} & \colhead{} \\
\cline{4-6} \colhead{} & \colhead{ } & \colhead{} & \colhead{} & \colhead{} & \colhead{} & \colhead{} & \colhead{} & \colhead{Optical} & \\
\colhead{2MASS Designation\tablenotemark{a}} &  \colhead{2MUCD} & \colhead{Other Name} &
\colhead{$J$} & \colhead{$J-H$} & \colhead{$J-K_S$} & \colhead{Obs. Date (UT)} & \colhead{Telescope}
& \colhead{Spectral Type\tablenotemark{b}}& \colhead{$M_J$}& \colhead{$d$ (pc)}& \colhead{Other Refs.\tablenotemark{c}}
}
\startdata
00054844$-$2157196	&	20005	&	LEHPM 162		&	13.274	&	0.657	&	1.073	&	2003 Nov 10	&	CT 1.5m	&	M9:		&	11.47$\pm$0.28	&	23.0$\pm$3.0	&	1, 2, 3 \\%Reyle04, Lodieu05, Kendall07		\\	
00065794$-$6436542	&	20007	&	\nodata			&	13.385	&	0.721	&	1.216	&	2003 Nov 09	&	CT 1.5m	&	M9:		&	11.47$\pm$0.28	&	24.2$\pm$3.1	&			\\	
00085931+2911521	&	20009	&	\nodata			&	13.827	&	0.700	&	1.137	&	2003 Jul 09	&	KP 4m	&	M8.5	&	11.32$\pm$0.15	&	31.7$\pm$2.3	&			\\	
00165953$-$4056541	&	20013	&	\nodata			&	15.316	&	1.110	&	1.884	&	2005 Aug 12	&	GS		&	L3.5\tablenotemark{d}	&	12.88$\pm$0.21	&	30.7$\pm$3.1	&	4\\%Davy. lithium.		\\	
00184613$-$6356122	&	20016	&	\nodata			&	15.224	&	0.995	&	1.613	&	2005 Aug 19	&	GS		&	L2		&	12.31$\pm$0.17	&	38.3$\pm$3.1	&			\\	
00285545$-$1927165	&	20022	&	\nodata			&	14.191	&	0.860	&	1.346	&	2004 Aug 09	&	CT 4m	&	L0:		&	11.73$\pm$0.26	&	31.1$\pm$3.8	&			\\	
00315477+0649463	&	20024	&	LSPM J0031+0649	&	12.820	&	0.636	&	1.102	&	2003 Oct 12	&	KP 2.1m	&	M7		&	10.73$\pm$0.25	&	26.1$\pm$3.0	&			\\	
00320509+0219017	&	20025	&	\nodata			&	14.324	&	0.938	&	1.522	&	2003 Jul 09	&	KP 4m	&	L1.5	&	12.15$\pm$0.15	&	27.2$\pm$2.0	&	5\\
\enddata
\end{deluxetable}

\begin{deluxetable}{lllllll}
\tablewidth{0pt}
\tablenum{5}
\tabletypesize{\small}
\tablecaption{Spectroscopically Confirmed Giants\label{tab:giants}}
\tablehead{\multicolumn{4}{c}{2MASS}  & \colhead{} & \colhead{} & \colhead{} \\
\cline{1-4} & \colhead{} & \colhead{} & \colhead{} & \colhead{} & \colhead{} & \colhead{Giant}\\
\colhead{Designation\tablenotemark{a}} & \colhead{$J$} & \colhead{$J-H$} &
\colhead{$J-K_S$} & \colhead{Date (UT)} & \colhead{Telescope} & \colhead{Spectral Class}  }
\startdata
00253141+8637440	&	9.989	&	0.841	&	1.306	&	2003 Oct 09	&	KP 2.1m	&	M pec		\\	%pec.	\\
00454220+4611578	&	9.570	&	0.835	&	1.266	&	2003 Oct 09	&	KP 2.1m	&	M8e 	\\%	Na and K in emission but no Halpha.	\\
02350658$-$5935521	&	9.798	&	0.722	&	1.156	&	2003 Nov 07	&	CT 1.5m	&	M7 			\\
02544898+1939372	&	13.823	&	0.856	&	1.343	&	2004 Feb 11	&	KP 4m	&	K5 		\\ %close to Pleiades	\\
05030162+6916324	&	9.271	&	0.806	&	1.221	&	2003 Mar 15	&	KP 2.1m	&	M5 			\\
05065349$-$0617123	&	11.160	&	1.245	&	2.026	&	2003 Mar 15	&	KP 2.1m	&	K4e 	\\	%emission.	\\
05293582$-$0629229	&	11.475	&	0.994	&	1.657	&	2003 Mar 15	&	KP 2.1m	&	K4e 	\\	%emission	\\
05302872$-$0928202	&	10.655	&	1.871	&	2.836	&	2003 Mar 15	&	KP 2.1m	&	M4e 	\\
\enddata

\tablenotetext{a}{The sexagesimal right ascension and declination suffix of the full 2MASS All-Sky Data Release designation 
(2MASS Jhhmmss[.]$\pm$ssddmmss[.]s) is listed for each object.
The coordinates are given for the J2000.0 equinox.}

\end{deluxetable}

\begin{deluxetable}{llllll}
\tablewidth{0pt}
\tablenum{6}
\tablecaption{Spectroscopically Confirmed Carbon Stars\label{tab:carbon}}
\tablehead{\multicolumn{3}{c}{2MASS} & \colhead{} & \colhead{} \\
\cline{1-4}\\
\colhead{Designation\tablenotemark{a}} & \colhead{$J$} & \colhead{$J-H$} &
\colhead{$J-K_S$} & \colhead{Obs. Date (UT)} & \colhead{Telescope}}
\startdata
01405432$-$7208509	&	12.205	&	1.125	&	1.890	&	2003 Nov 09	&	CT 1.5m	\\
02000890+4137474	&	10.867	&	1.318	&	2.211	&	2003 Oct 09	&	KP 2.1m	\\
03463857+7547165	&	10.378	&	1.003	&	1.541	&	2003 Oct 09	&	KP 2.1m	\\
03595596+0919044	&	12.757	&	1.528	&	2.643	&	2003 Nov 07	&	CT 1.5m	\\
05283374$-$1651445	&	9.312	&	1.011	&	1.500	&	2003 Mar 15	&	KP 2.1m	\\
07342392+2719115	&	9.273	&	1.085	&	1.734	&	2003 Mar 13	&	KP 2.1m	\\
07522490+0433586	&	9.233	&	0.910	&	1.517	&	2003 Mar 13	&	KP 2.1m	\\
08491096$-$0721442	&	10.512	&	0.988	&	1.448	&	2003 Mar 14	&	KP 2.1m	\\
\enddata

\tablenotetext{a}{The sexagesimal right ascension and declination suffix of the full 2MASS All-Sky Data Release designation 
(2MASS Jhhmmss[.]$\pm$ssddmmss[.]s) is listed for each object.
The coordinates are given for the J2000.0 equinox.}

\end{deluxetable}

\begin{deluxetable}{llcccccccrcc}
\tabletypesize{\tiny}
\tablenum{7}
\rotate
\tablewidth{0pt}
\tablecaption{L Dwarf Spectral Standards}
\tablehead{
\colhead{2MASS Designation\tablenotemark{a}} & \colhead{Other Name} & \colhead{Spectral Type} & \colhead{2MASS J} & 
\colhead{$\mu$ ($\arcsec$ yr$^{-1}$)} & \colhead{PA ($\degr$)} &  
\colhead{CrH-a} & \colhead{Rb-b/TiO-b} & \colhead{Cs-a/VO-b} & \colhead{Color-d}  & \colhead{Multiple?} & \colhead{Refs.}
}
\startdata
\cutinhead{Secondary Standards}
2MASS J17312974+2721233   & U20744  LSPM J1731+2721      & L0 & 12.094 & 0.269$\pm$0.008   & 200$\pm$2      & 1.18 & 0.59 & 0.71 & 5.68    & N & 1--4\\
2MASS J06023045+3910592   & 50010   LSR J0602+3910       & L1 & 12.300 & 0.526$\pm$0.006   & 164.1$\pm$0.8  & 1.25 & 0.82 & 0.85 & 6.79    & ? & 2,5\\
2MASS J08472872-1532372 & 10764   \nodata              & L2 & 13.513 & 0.274$\pm$0.039   & 146$\pm$10     & 1.48 & 1.01 & 0.98 & 8.52    & N & 3,4,6\\
2MASS J15065441+1321060   & U11291  \nodata              & L3 & 13.365 & 1.093$\pm$0.019   & 271$\pm$1      & 1.75 & 1.18 & 1.13 & 7.80    & N & 3,7--9\\
2MASS J05002100+0330501   & u20197  \nodata              & L4 & 13.669 & 0.362$\pm$0.043   & 182$\pm$9      & 1.95 & 1.26 & 1.19 & \nodata & N & 1,3,4\\
%2MASS J06523073+4710348   & U10601  \nodata              & L5 & 13.511 & 0.135$\pm$0.045   & 336$\pm$13     & 2.10 & 1.70 & 1.49 & 14.73   & N & 3,4,6,\\
2MASS J15074769-1627386  & U11296 \nodata              & L5 & 12.830 & 0.9031$\pm$0.0005 & 190$\pm$0.1   & 2.00 & 1.46 & 1.39 & 13.19 & N & 4,9--11 \\
2MASS J02550357-4700509 & 10158   DENIS-P J0255$-$4700 & L8 & 13.246 & 1.149$\pm$0.002   & 119.5$\pm$0.2  & 1.29 & 2.20 & 1.69 & 36.62   & N & 3,9,12--14\\ 
\cutinhead{Primary Standards from Kirkpatrick et al, 1999}
2MASS J03454316+2540233  & 20165    \nodata             & L0 & 13.997 & 0.1020$\pm$0.0003 & 249.6$\pm$0.2  & 1.16 & 0.66 & 0.78 & 6.65    & N & 8,11,15\\
\enddata
\tablerefs{
(1)~This Paper; (2)~L\'epine \& Shara, 2005; (3)~Schmidt \etal, 2007; (4)~Reid \etal, 2006a; (5)~Salim \etal, 2003; (6)~Cruz \etal, 2003; (7)~Gizis \etal, 2000; (8)~Gizis \etal, 2003; (9)~Reid \etal, 2007; (10)~Reid \etal, 2000;  
(11)~Dahn \etal, 2002; (12)~\pedant \etal, 1999; (13)~Deacon \etal, 2005; 
(14)~Costa \etal, 2005;(15)~Kirkpatrick \etal, 1999; (16)~Reid \etal, 2001;
(17)~Ruiz \etal, 1997; (18)~liu \& Legget, 2005;(19)~Koerner \etal, 1999; 
(20)~\pedant \etal, 2006; (21)~Delfosse \etal, 1997; (22)~Vrba \etal, 2004.
}
%\tablecomments{}
\tablenotetext{a}{The sexagesimal right ascension and declination suffix of the full 2MASS All-Sky Data Release designation 
(2MASS Jhhmmss[.]$\pm$ssddmmss[.]s) is listed for each object.
The coordinates are given for the J2000.0 equinox; the units of right ascension are hours, minutes, and seconds; 
and units of declination are degrees, arcminutes, and arcseconds.}
\label{tab:newstds}
\end{deluxetable}

\begin{deluxetable}{rrclllc}
\tablecolumns{7}
\tablenum{8}
\tablewidth{0pt}
\tablecaption{L dwarfs with lithium absorption}
 \tablehead{\colhead{2MASS J} & \colhead {2MUCD} & \colhead {Sp. Type} &
\colhead {E.W. (\AA)} & \colhead{d (pc.)} & \colhead{M$_J$} & \colhead {Notes} } 
\startdata
 \cutinhead{L dwarfs from this paper} 
00165953-4056541 & 20013 & L3.5 & 5.1 & 30.7 & 12.9 & Gemini\\
03231002-4631237 & 20157 & (L0) & 3.5 & 54 & 11.7 & Gemini \\ 
03552337+1133437 & 20171 & (L5) & 7.0 & 18.8 & 12.7 & KPNO/Gemini\\
05012406-0010452 & 20198 & (L4) & 9.0 & 29.0 & 12.7 & KPNO/Gemini \\
06322402-5010349 & 20248 & L3 & 7.6 & 29.5 & 12.7 & Gemini \\
09054654+5623117 & 20329 & L5 & 9.8 & 23.3 & 13.6 & Gemini \\
19223062+6610194 & 20812 & L1 & 3.7 & 32.7 & 12.0 & Gemini \\
23174712-4838501 & 20994 & L4 & 12.2 & 25.8 & 13.1 & Gemini \\
 \cutinhead{L dwarfs from Papers V \& IX} 
0025036+475919 & 13016 & L4: & 8.2 & 32 & 13.1 & Gemini, binary, R \\
0310140-275645 & 10170 & L5 & 10.4 & 28.3 & 13.6 & KPNO/Gemini, IX \\
0326422-210205 & 10184 & L4 & 11.9 & 40.1 & 13.1 & CTIO/Gemini, IX \\
0421072-630602 & 10268 & (L5) & 7.7 & 31.1 & 13.1 & Gemini, IX \\
0512063-294954 & 10372 & L4 & 7.5 & 30.5 & 13.1 & CTIO/Gemini, V \\
0652307+471034 & 10601 & L4.5 & 14.7 & 11.1 & 13.3 & KPNO/Gemini, IX \\
1615425+495321 & 11538 & (L4) & 12 & 53 & 13.1 & Gemini, IX \\
\enddata
\tablecomments{References: \\
V, Cruz \etal (2003), Paper V; IX, Cruz \etal (2006), Paper IX; R, Reid \etal (2006). }
\end{deluxetable}

\begin{deluxetable}{rcccrrrccl}
\tabletypesize{\scriptsize}
\tablecolumns{10}
\tablenum{9}
\tablewidth{0pt}
\tablecaption{L dwarf systems within 20 parsecs of the Sun}
\tablehead{\colhead {N} & \colhead{2MASS J} & \colhead {2MUCD} & \colhead{SpT} & 
\colhead{J} & \colhead{(J-H)} & \colhead{(H-K$_S$)} &
\colhead{d (pc.)} & \colhead{src.} & \colhead{Notes} } 
\startdata
1$^*$ & 00043484-4044058Ba & 20004A & L4.5 & 13.82 & 0.95 & 0.65 & 9.6 & trig & LHS 102Ba, 1, 2 \\
 & 00043484-4044058Bb & 20004B & L4.5 & 13.90 & 0.95 & 0.65 & 9.6 & trig & LHS 102Bb, 1, 2 \\
2$^*$  & 00361617+1821104 & 20029 & L3.5 & 12.47 & 0.88 & 0.53 & 8.76 & trig & 3,4 \\
3$^*$  & 00452143+1634446 & 20037 & (L2) & 13.06 & 1.00 & 0.69 & 14 & sp & 9, H$\alpha$ 14\AA \\
4$^*$  & 01075242+0041563 & 20052 & L8 & 15.82 & 1.31 & 0.80 & 15.6 & trig & 5, 6\\
5 & 01282664-5545343 & 20068 & L2 & 13.78 & 0.86 & 0.58 & 19.6 & sp & \\
6 & 0144353-071614 & 10088 & L5 & 14.19 & 1.19 & 0.72 & 13.4 & sp & 7,8\\
7$^*$  & 01550354+0950003 & 20083 & L5 & 14.83 & 1.06 & 0.63 & 17.95 & sp & 9 \\
\enddata
\tablecomments{Sources from the 2MU2 sample have 2MUCD 5-digit numbers 1xxxx; sources from the 2MUA sample have 2MUCD numbers 2xxxx; \\
$^*$ indicates that the L dwarf has HST NICMOS observations (Reid \etal, 2007; 2008). \\
References: \\
1, EROS collaboration, 1999; 2, Golimowski et al, 2004; 3, Reid \etal, 2000; 4, Dahn \etal, 2002; 5, Hawley \etal, 2002; 6, Vrba \etal, 2004; 7, Cruz \etal, 2003; 8, Liebert \etal, 2003; 9, this paper ; 10, Delfosse et al, 1997; 11, Koerner \etal, 1999; 12, Bouy et al, 2005; 13, Bartlett et al, 2007; 14, \pedant \etal, 1999; 15, Costa et al, 2006;  16, Looper et al, subm.; 17, Cruz et al, 2006 18, Geballe et al, 2002; 19, Burgasser \etal, 2005; 20, Salim et al, 2003; 21, Tinney, Burgasser \& Kirkpatrick, 2003; 22, Reid et al, 2006; 23, Reid \etal, 2001; 24, Kirkpatrick et al, 1999;  25, Scholz \etal, 2002; 26, Gizis \etal, 2000; 27, Gizis, 2002; 28, Fan \etal, 2000; 29, Ruiz \etal, 1997; 30, Liu \& Leggett, 2005; 31, Kendall \etal, 2004; 32, Burgasser \etal, 2007; 33, Kendall \etal, 2007; 34, Deacon \& Hambly, 2007; 35, Kirkpatrick \etal, 2000; 36, Reid \etal, 2006; 37, Folkes \etal, 2007; 38, Burgasser \etal, 2008; 39, Phan-Bao \etal, 2008; 40, Burgasser, 2007b\\}
\end{deluxetable}


\begin{thebibliography}

\bibitem[Allen]{al05} Allen, P.R., Koerner, D.W., Reid, I.N., Trilling, D.E. 2005, \apj, 625, 385

\bibitem[Baraffe et al.(1998)]{1998A&A...337..403B} Baraffe, I., Chabrier, G., Allard, F., \& Hauschildt, P.~H.\ 1998, \aap, 337, 403 

\bibitem[{{Bartlett}(2007)}]{Bartlett07} {Bartlett}, J.~L. 2007, PhD thesis, University of Virginia

\bibitem[{{Basri} {et~al.}(2000){Basri}, {Mohanty}, {Allard}, {Hauschildt}, {Delfosse}, {Mart{\'{\i}}n}, {Forveille}, \& {Goldman}}]{Basri00} {Basri}, G., {Mohanty}, S., {Allard}, F., {Hauschildt}, P.~H., {Delfosse}, X., {Mart{\'{\i}}n}, E.~L., {Forveille}, T., \& {Goldman}, B. 2000, \apj, 538, 363

\bibitem[Bessell, 1990] {bs90} Bessell, M.S. 1990, \aaps, 83, 357

\bibitem[{{Bouy} {et~al.}(2003){Bouy}, {Brandner}, {Mart{\'{\i}}n}, {Delfosse},
  {Allard}, \& {Basri}}]{Bouy03} {Bouy}, H., {Brandner}, W., {Mart{\'{\i}}n}, E.~L., {Delfosse}, X., {Allard}, F., \& {Basri}, G. 2003, \aj, 126, 1526

\bibitem[Bouy et al.(2005)]{2005AJ....129..511B} Bouy, H., Mart{\'{\i}}n, E.~L., Brandner, W., \& Bouvier, J.\ 2005, \aj, 129, 511 

\bibitem[Burgasser et al.(2002)]{2002ApJ...564..421B} Burgasser, A.~J., et 
al.\ 2002, \apj, 564, 421 

\bibitem[Burgasser(2007)]{2007ApJ...659..655B} Burgasser, A.~J.\ 2007a, \apj, 659, 655 

\bibitem[Burgasser(2007)]{2007AJ....134.1330B} Burgasser, A.~J.\ 2007b, \aj, 134, 1330 

\bibitem[Burgasser et al, 2008] {burg08} Burgasser, A.J., Looper, D.L., Kirkpatrick, J.D., Cruz, K.L., Swift, B.J., 2008, \apj, 674, 451

\bibitem[{{Burgasser} \& {McElwain}(2006)}]{Burgasser06_2200} {Burgasser}, A.~J., \& {McElwain}, M.~W. 2006, \aj, 131, 1007

\bibitem[Burgasser et al.(2005)]{2005ApJ...634L.177B} Burgasser, A.~J., Reid, I.~N., Leggett, S.~K., Kirkpatrick, J.~D., Liebert, J., \& Burrows, A.\ 2005, \apjl, 634, L177 

\bibitem[Burgasser et al.(2006)]{2006ApJ...637.1067B} Burgasser, A.~J., 
Geballe, T.~R., Leggett, S.~K., Kirkpatrick, J.~D., \& Golimowski, D.~A.\ 
2006, \apj, 637, 1067 

\bibitem[Burgasser et al.(2007)]{me06ppv}Burgasser, A.\ J., Reid, I.\ N., Siegler, N., Close, L.\ M., Allen, P., Lowrance, P.\ J., \& Gizis, J.\ E. 2007, in Planets and Protostars V, eds.\ B.\ Reipurth, D.\ Jewitt and K.\ Keil (Univ.\ Arizona Press: Tucson), 427

\bibitem[Burrows et al.(2001)]{2001RvMP...73..719B} Burrows, A., Hubbard, W.~B., Lunine, J.~I., \& Liebert, J.\ 2001, Reviews of Modern Physics, 73, 719 

\bibitem[Chabrier \& Baraffe(1997)]{1997A&A...327.1039C} Chabrier, G., \& Baraffe, I.\ 1997, \aap, 327, 1039 

\bibitem[{{Costa} {et~al.}(2005){Costa}, {M{\'e}ndez}, {Jao}, {Henry},  {Subasavage}, {Brown}, {Ianna}, \& {Bartlett}}]{Costa05} {Costa}, E., {M{\'e}ndez}, R.~A., {Jao}, W.-C., {Henry}, T.~J., {Subasavage}, J.~P., {Brown}, M.~A., {Ianna}, P.~A., \& {Bartlett}, J. 2005, \aj, 130, 337

\bibitem[Costa et al.(2006)]{2006AJ....132.1234C} Costa, E., M{\'e}ndez, R.~A., Jao, W.-C., Henry, T.~J., Subasavage, J.~P., \& Ianna, P.~A.\ 2006, \aj, 132, 1234 

\bibitem[{{Crifo} {et~al.}(2005){Crifo}, {Phan-Bao}, {Delfosse}, {Forveille}, {Guibert}, {Mart{\'{\i}}n}, \& {Reyl{\'e}}}]{NN6} {Crifo}, F., {Phan-Bao}, N., {Delfosse}, X., {Forveille}, T., {Guibert}, J., {Mart{\'{\i}}n}, E.~L., \& {Reyl{\'e}}, C. 2005, \aap, 441, 653

\bibitem[Cruz et al.(2003)]{2003AJ....126.2421C} Cruz, K.~L., Reid, I.~N., 
Liebert, J., Kirkpatrick, J.~D., \& Lowrance, P.~J.\ 2003, \aj, 126, 2421
[Paper V]

\bibitem[{{Cruz} {et~al.}(2007){Cruz}, {Reid}, {Kirkpatrick}, {Burgasser}, {Liebert}, {Solomon}, {Schmidt}, {Allen}, {Hawley}, \& {Covey}}]{Cruz07} {Cruz}, K.~L., {Reid}, I.~N., {Kirkpatrick}, J.~D., {Burgasser}, A.~J., {Liebert}, J., {Solomon}, A.~R., {Schmidt}, S.~J., {Allen}, P.~R., {Hawley}, S.~L., \& {Covey}, K.~R. 2007, \aj, 133, 439 [Paper IX]

\bibitem [Dahn {\sl et al.}, 2002] {da} Dahn, C.C. {\sl et al.} 2002, \aj, 124, 1170

\bibitem[Dame et al.(1987)]{Dame} Dame, T.~M.~et al.\ 1987, \apj, 322, 706

\bibitem[{{Deacon} {et~al.}(2005){Deacon}, {Hambly}, \& {Cooke}}]{Deacon05}
{Deacon}, N.~R., {Hambly}, N.~C., \& {Cooke}, J.~A. 2005, \aap, 435, 363

\bibitem[Deacon \& Hambly(2007)]{2007A&A...468..163D} Deacon, N.~R., \& Hambly, N.~C.\ 2007, \aap, 468, 163 

\bibitem[Delfosse et al.(1997)]{1997A&A...327L..25D} Delfosse, X., et al.\ 
1997, \aap, 327, L25 

\bibitem[Delfosse et al.(1999)]{1999A&A...341L..63D} Delfosse, X., Forveille, T., Mayor, M., Burnet, M., \& Perrier, C.\ 1999a, \aap, 341, L63 

\bibitem[Delfosse et al.(1999)]{1999A&A...344..897D} Delfosse, X., 
Forveille, T., Beuzit, J.-L., Udry, S., Mayor, M., \& Perrier, C.\ 1999b, 
\aap, 344, 897 

\bibitem[{{Delfosse} {et~al.}(2001){Delfosse}, {Forveille}, {Mart{\'{\i}}n},
  {Guibert}, {Borsenberger}, {Crifo}, {Alard}, {Epchtein}, {Fouqu{\'e}},
  {Simon}, \& {Tajahmady}}]{Delfosse01}
{Delfosse}, X., {Forveille}, T., {Mart{\'{\i}}n}, E.~L., {Guibert}, J.,
  {Borsenberger}, J., {Crifo}, F., {Alard}, C., {Epchtein}, N., {Fouqu{\'e}},
  P., {Simon}, G., \& {Tajahmady}, F. 2001, \aap, 366, L13

\bibitem[{{Dobbie} {et~al.}(2002){Dobbie}, {Kenyon}, {Jameson}, {Hodgkin},
  {Hambly}, \& {Hawkins}}]{Dobbie02}
{Dobbie}, P.~D., {Kenyon}, F., {Jameson}, R.~F., {Hodgkin}, S.~T., {Hambly},
  N.~C., \& {Hawkins}, M.~R.~S. 2002, \mnras, 329, 543

\bibitem[Duquennoy \& Mayor(1991)]{1991A&A...248..485D} Duquennoy, A.~\& 
Mayor, M.\ 1991, \aap, 248, 485 

\bibitem[Dutra \& Bica(2002)]{CDN} Dutra, C.~M.~\& Bica, E.\ 2002, \aap, 383, 631

\bibitem[Epchtein et al.(1994)]{1994Ap&SS.217....3E} Epchtein, N., et al.\ 
1994, \apss, 217, 3 

\bibitem[EROS Collaboration et al.(1999)]{1999A&A...351L...5E} EROS Collaboration, et al.\ 1999, \aap, 351, L5 

\bibitem[Fan et al.(2000)]{2000AJ....119..928F} Fan, X., et al.\ 2000, \aj, 
119, 928 

\bibitem[Folkes et al.(2007)]{2007MNRAS.378..901F} Folkes, S.~L., Pinfield,  D.~J., Kendall, T.~R., \& Jones, H.~R.~A.\ 2007, \mnras, 378, 901 

\bibitem[Geballe et al.(2002)]{2002ApJ...564..466G} Geballe, T.~R., et al.\ 
2002, \apj, 564, 466 

\bibitem[Gizis(2002)]{2002ApJ...575..484G} Gizis, J.~E.\ 2002, \apj, 575, 
484 

\bibitem[Gizis et al.(2000)]{2000AJ....120.1085G} Gizis, J.~E., Monet, 
D.~G., Reid, I.~N., Kirkpatrick, J.~D., Liebert, J., \& Williams, R.~J.\ 
2000, \aj, 120, 1085 

\bibitem[{{Gizis} \& {Reid}(1997)}]{GR97} {Gizis}, J.~E., \& {Reid}, I.~N. 1997, \pasp, 109, 849

\bibitem[{{Gizis} {et~al.}(1999){Gizis}, {Reid}, \& {Monet}}]{Gizis99}
{Gizis}, J.~E., {Reid}, I.~N., \& {Monet}, D.~G. 1999, \aj, 118, 997

\bibitem[{{Gizis} {et~al.}(2003){Gizis}, {Reid}, {Knapp}, {Liebert},
  {Kirkpatrick}, {Koerner}, \& {Burgasser}}]{Gizis03}
{Gizis}, J.~E., {Reid}, I.~N., {Knapp}, G.~R., {Liebert}, J., {Kirkpatrick},
  J.~D., {Koerner}, D.~W., \& {Burgasser}, A.~J. 2003, \aj, 125, 3302

\bibitem [Gliese & Jahreiss] {gj91}  Gliese, W., Jahrei{\ss},  1991, 
Preliminary Version of the Third Catalogue of Nearby Stars, (CNS3)

\bibitem[Golimowski et al.(2004)]{2004AJ....128.1733G} Golimowski, D.~A., et al.\ 2004, \aj, 128, 1733 

\bibitem[Hamuy et al.(1994)]{1994PASP..106..566H} Hamuy, M., Suntzeff, 
N.~B., Heathcote, S.~R., Walker, A.~R., Gigoux, P., \& Phillips, M.~M.\ 
1994, \pasp, 106, 566 

\bibitem[{{Hawkins} \& {Bessell}(1988)}]{LHS2065} {Hawkins}, M.~R.~S., \& {Bessell}, M.~S. 1988, \mnras, 234, 177

\bibitem[Hawley et al.(2002)]{2002AJ....123.3409H} Hawley, S.~L., et al.\ 
2002, \aj, 123, 3409 

\bibitem[{{Henry} {et~al.}(2004){Henry}, {Subasavage}, {Brown}, {Beaulieu},
  {Jao}, \& {Hambly}}]{Henry04}
{Henry}, T.~J., {Subasavage}, J.~P., {Brown}, M.~A., {Beaulieu}, T.~D., {Jao},
  W.-C., \& {Hambly}, N.~C. 2004, \aj, 128, 2460

\bibitem[{{Hook} {et~al.}(2004){Hook}, {J{\o}rgensen}, {Allington-Smith},
  {Davies}, {Metcalfe}, {Murowinski}, \& {Crampton}}]{GMOS}
{Hook}, I.~M., {J{\o}rgensen}, I., {Allington-Smith}, J.~R., {Davies}, R.~L.,
  {Metcalfe}, N., {Murowinski}, R.~G., \& {Crampton}, D. 2004, \pasp, 116, 425

\bibitem[{{Ianna} \& {Fredrick}(1995)}]{Ianna95}
{Ianna}, P.~A., \& {Fredrick}, L.~W. 1995, \apjl, 441, L47

\bibitem[{Jameson {et~al.}(2007)Jameson, Casewell, Bannister, Lodieu,
  Keresztes, Dobbie, \& Hodgkin}]{Jameson07}
Jameson, R.~F., Casewell, S.~L., Bannister, N.~P., Lodieu, N., Keresztes, K.,
  Dobbie, P.~D., \& Hodgkin, S.~T. 2007

\bibitem[{{Jao} {et~al.}(2005){Jao}, {Henry}, {Subasavage}, {Brown}, {Ianna},
  {Bartlett}, {Costa}, \& {M{\'e}ndez}}]{Jao05} {Jao}, W.-C., {Henry}, T.~J., {Subasavage}, J.~P., {Brown}, M.~A., {Ianna},  P.~A., {Bartlett}, J.~L., {Costa}, E., \& {M{\'e}ndez}, R.~A. 2005, \aj, 129, 1954

\bibitem[Kendall et al.(2004)]{2004A&A...416L..17K} Kendall, T.~R., Delfosse, X., Mart{\'{\i}}n, E.~L., \& Forveille, T.\ 2004, \aap, 416, L17 

\bibitem[Kendall et al.(2007)]{2007MNRAS.374..445K} Kendall, T.~R., Jones, H.~R.~A., Pinfield, D.~J., Pokorny, R.~S., Folkes, S., Weights, D., Jenkins, J.~S., \& Mauron, N.\ 2007, \mnras, 374, 445 

\bibitem[Kirkpatrick et al.(1999)]{K99} Kirkpatrick, J.~D.~et al.\ 1999, \apj, 519, 802 

\bibitem[Kirkpatrick et al.(2000)]{2000AJ....120..447K} Kirkpatrick, J.~D., 
et al.\ 2000, \aj, 120, 447 

\bibitem[Kirkpatrick et al.(2006)]{2006ApJ...639.1120K} Kirkpatrick, J.~D., 
Barman, T.~S., Burgasser, A.~J., McGovern, M.~R., McLean, I.~S., Tinney, 
C.~G., \& Lowrance, P.~J.\ 2006, \apj, 639, 1120 

\bibitem[{{Kirkpatrick} {et~al.}(1997){Kirkpatrick}, {Henry}, \&
  {Irwin}}]{KHI97}
{Kirkpatrick}, J.~D., {Henry}, T.~J., \& {Irwin}, M.~J. 1997, \aj, 113, 1421

\bibitem[{{Kirkpatrick} {et~al.}(1991){Kirkpatrick}, {Henry}, \&
  {McCarthy}}]{KHM91}
{Kirkpatrick}, J.~D., {Henry}, T.~J., \& {McCarthy}, D.~W. 1991, \apjs, 77, 417

\bibitem[{{Kirkpatrick} {et~al.}(1995){Kirkpatrick}, {Henry}, \&
  {Simons}}]{K95}
{Kirkpatrick}, J.~D., {Henry}, T.~J., \& {Simons}, D.~A. 1995, \aj, 109, 797

\bibitem[Kirkpatrick et al, 2008] {k2008} Kirkpatrick, J.~D., Looper, D., 2008, \aj, subm.

\bibitem[Knapp et al.(2004)]{2004AJ....127.3553K} Knapp, G.~R., et al.\ 
2004, \aj, 127, 3553 

\bibitem[Koerner et al.(1999)]{1999ApJ...526L..25K} Koerner, D.~W., 
Kirkpatrick, J.~D., McElwain, M.~W., \& Bonaventura, N.~R.\ 1999, \apjl, 
526, L25 

\bibitem[{{Law} {et~al.}(2006){Law}, {Hodgkin}, \& {Mackay}}]{Law06}
{Law}, N.~M., {Hodgkin}, S.~T., \& {Mackay}, C.~D. 2006, \mnras, 368, 1917

\bibitem[L{\'e}pine(2008)]{2008AJ....135.2177L} L{\'e}pine, S.\ 2008, \aj, 
135, 2177 

\bibitem[L{\'e}pine \& Shara(2005)]{2005AJ....129.1483L} L{\'e}pine, S., \& Shara, M.~M.\ 2005, \aj, 129, 1483 

\bibitem[Liebert et al.(2003)]{2003AJ....125..343L} Liebert, J., Kirkpatrick, J.~D., Cruz, K.~L., Reid, I.~N., Burgasser, A., Tinney, C.~G., \& Gizis, J.~E.\ 2003, \aj, 125, 343 

\bibitem[Liu \& Leggett(2005)]{2005ApJ...634..616L} Liu, M.~C., \& Leggett, S.~K.\ 2005, \apj, 634, 616 

\bibitem[{{Lodieu} {et~al.}(2002){Lodieu}, {Scholz}, \& {McCaughrean}}] {Lodieu02} {Lodieu}, N., {Scholz}, R.-D., \& {McCaughrean}, M.~J. 2002, \aap, 389, L20

\bibitem[{{Lodieu} {et~al.}(2005){Lodieu}, {Scholz}, {McCaughrean}, {Ibata},
  {Irwin}, \& {Zinnecker}}]{Lodieu05}
{Lodieu}, N., {Scholz}, R.-D., {McCaughrean}, M.~J., {Ibata}, R., {Irwin}, M.,
  \& {Zinnecker}, H. 2005, \aap, 440, 1061

\bibitem[Looper et al] {loop} Looper, D.L., Kirkpatrick, J.D. et al, 2008, \apj, subm.

\bibitem[Lowrance et al, 2003] {low03} Lowrance, P.J., Kirkpatrick, J.D., Reid, I.N., Cruz, K.L. 2003, \apj, 584, L95

\bibitem[Luyten(1980)]{1980nlca.book.....L} Luyten, W.\ 1980, The New Luyten Two-Tenths Catalogue )NLTT), Minneapolis: University of Minnesota, 1980,  

\bibitem[Marley et al.(2002)]{2002ApJ...568..335M} Marley, M.~S., Seager, S., Saumon, D., Lodders, K., Ackerman, A.~S., Freedman, R.~S., \& Fan, X.\ 2002, \apj, 568, 335 

\bibitem[{{Mart{\'{\i}}n} {et~al.}(2006){Mart{\'{\i}}n}, {Brandner}, {Bouy},
  {Basri}, {Davis}, {Deshpande}, \& {Montgomery}}]{Martin06}
{Mart{\'{\i}}n}, E.~L., {Brandner}, W., {Bouy}, H., {Basri}, G., {Davis}, J.,
  {Deshpande}, R., \& {Montgomery}, M.~M. 2006, \aap, 456, 253

\bibitem[Mart{\'{\i}}n et al.(1999)]{1999AJ....118.2466M} Mart{\'{\i}}n, 
E.~L., Delfosse, X., Basri, G., Goldman, B., Forveille, T., \& Zapatero 
Osorio, M.~R.\ 1999, \aj, 118, 2466 

\bibitem[{{Massey} \& {Gronwall}(1990)}]{Massey90} {Massey}, P., \& {Gronwall}, C. 1990, \apj, 358, 344

\bibitem[{{Massey} {et~al.}(1988){Massey}, {Strobel}, {Barnes}, \&
  {Anderson}}]{Massey88} {Massey}, P., {Strobel}, K., {Barnes}, J.~V., \& {Anderson}, E. 1988, \apj, 328, 315

\bibitem[{{McElwain} \& {Burgasser}(2006)}]{McElwain06}
{McElwain}, M.~W., \& {Burgasser}, A.~J. 2006, \aj, 132, 2074

\bibitem[{{McLean} {et~al.}(2003){McLean}, {McGovern}, {Burgasser},
  {Kirkpatrick}, {Prato}, \& {Kim}}]{McLean03}
{McLean}, I.~S., {McGovern}, M.~R., {Burgasser}, A.~J., {Kirkpatrick}, J.~D.,
  {Prato}, L., \& {Kim}, S.~S. 2003, \apj, 596, 561

\bibitem[{{Monet} {et~al.}(1992){Monet}, {Dahn}, {Vrba}, {Harris}, {Pier},
  {Luginbuhl}, \& {Ables}}]{Monet92}
{Monet}, D.~G., {Dahn}, C.~C., {Vrba}, F.~J., {Harris}, H.~C., {Pier}, J.~R.,
  {Luginbuhl}, C.~B., \& {Ables}, H.~D. 1992, \aj, 103, 638

\bibitem[Nakajima et al.(1995)]{1995Natur.378..463N} Nakajima, T., 
Oppenheimer, B.~R., Kulkarni, S.~R., Golimowski, D.~A., Matthews, K., \& 
Durrance, S.~T.\ 1995, \nat, 378, 463 

\bibitem[{{Neuh{\"a}user} {et~al.}(2002){Neuh{\"a}user}, {Guenther}, {Alves},
  {Grosso}, {Leinert}, {Ratzka}, {Ott}, {Mugrauer}, {Comer{\'o}n}, {Brandner},
  \& {Eckart}}]{Neuhauser02}
Neuh{\"a}user, R. et al., 2002, Astronomische Nachrichten, 323, 447

\bibitem[Oke \& Gunn(1983)]{1983ApJ...266..713O} Oke, J.~B.~\& Gunn, J.~E.\ 
1983, \apj, 266, 713 

\bibitem[{{Perryman} \& {ESA}(1997)}]{Hipparcos}
{Perryman}, M.~A.~C., \& {ESA}. 1997, {The HIPPARCOS and TYCHO catalogues.
  Astrometric and photometric star catalogues derived from the ESA HIPPARCOS
  Space Astrometry Mission} (The Hipparcos and Tycho catalogues.~Astrometric
  and photometric star catalogues derived from the ESA Hipparcos Space
  Astrometry Mission, Publisher: Noordwijk, Netherlands: ESA Publications
  Division, 1997, Series: ESA SP Series vol no: 1200, ISBN: 9290923997 (set))

\bibitem[Pinfield et al.(2006)]{2006MNRAS.368.1281P} Pinfield, D.~J., 
Jones, H.~R.~A., Lucas, P.~W., Kendall, T.~R., Folkes, S.~L., Day-Jones, 
A.~C., Chappelle, R.~J., \& Steele, I.~A.\ 2006, \mnras, 368, 1281 

\bibitem[Phan-Bao et al.(2001)]{2001A&A...380..590P} Phan-Bao, N.~et al.\ 2001, \aap, 380, 590 

\bibitem[Phan-Bao et al.(2008)]{2008PB} Phan-Bao, N.~et al.\ 2008, \mnras, in press

\bibitem[{{Phan-Bao} \& {Bessell}(2006)}]{PhanBao06}
{Phan-Bao}, N., \& {Bessell}, M.~S. 2006, \aap, 446, 515

\bibitem[{{Phan-Bao} {et~al.}(2003){Phan-Bao}, {Crifo}, {Delfosse},
  {Forveille}, {Guibert}, {Borsenberger}, {Epchtein}, {Fouqu{\'e}}, {Simon}, \&
  {Vetois}}]{NN5}
{Phan-Bao}, N., {Crifo}, F., {Delfosse}, X., {Forveille}, T., {Guibert}, J.,
  {Borsenberger}, J., {Epchtein}, N., {Fouqu{\'e}}, P., {Simon}, G., \&
  {Vetois}, J. 2003, \aap, 401, 959

\bibitem[{{Probst} \& {Liebert}(1983)}]{LHS2924}
{Probst}, R.~G., \& {Liebert}, J. 1983, \apj, 274, 245

\bibitem[Rebolo et al, 1992] {rmm} Rebolo, R., Mart{\'{\i}}n, E.L., Magazzu, A., 1992, \apj, 389, L83

\bibitem[Reid et al.(2004)]{2004AJ....128..463R} Reid, I.~N., et al.\ 2004, \aj, 128, 463 [Paper VIII]

\bibitem[Reid et al.(2007)]{2007AJ....133.2825R} Reid, I.~N., Cruz, K.~L.,
\& Allen, P.~R.\ 2007, \aj, 133, 2825 [Paper XI]

\bibitem [Reid et al, 2008] {paper} Reid, I.~N., Cruz, K.~L., Burgasser, A.~J.~\& Liu, M.~C. 2008, \aj, in press 

\bibitem[Reid \& Gizis (1997)] {rg97} Reid, I.~N.~\& Gizis, J.~E., 1997, 
\aj, 113, 2246

\bibitem[Reid et al.(2001)]{2001AJ....121..489R} Reid, I.~N., Gizis, J.~E., 
Kirkpatrick, J.~D., \& Koerner, D.~W.\ 2001, \aj, 121, 489 

\bibitem[{{Reid} \& {Gizis}(2005)}]{RG05} {Reid}, I.~N., \& {Gizis}, J.~E. 2005, \pasp, 117, 676

\bibitem[Reid et al.(2000)]{2000AJ....119..369R} Reid, I.~N., Kirkpatrick, J.~D., Gizis, J.~E., Dahn, C.~C., Monet, D.~G., Williams, R.~J., Liebert,  J., \& Burgasser, A.~J.\ 2000, \aj, 119, 369 

\bibitem[Reid et al.(2006)]{2006AJ....132..891R} Reid, I.~N., Lewitus, E., Allen, P.~R., Cruz, K.~L., \& Burgasser, A.~J.\ 2006a, \aj, 132, 891 

\bibitem[Reid et al.(2006)]{2006ApJ...639.1114R} Reid, I.~N., Lewitus, E., Burgasser, A.~J., \& Cruz, K.~L.\ 2006b, \apj, 639, 1114 

\bibitem[Reid et al.(2007)]{2007AJ....133.2825R} Reid, I.~N., Cruz, K.~L., \& Allen, P.~R.\ 2007, \aj, 133, 2825 
\bibitem[{{Reyl{\'e}} \& {Robin}(2004)}]{Reyle04}
{Reyl{\'e}}, C., \& {Robin}, A.~C. 2004, \aap, 421, 643

\bibitem[{{Reyl{\'e}} {et~al.}(2006){Reyl{\'e}}, {Scholz}, {Schultheis},
  {Robin}, \& {Irwin}}]{Reyle06}
{Reyl{\'e}}, C., {Scholz}, R.-D., {Schultheis}, M., {Robin}, A.~C., \& {Irwin},
  M. 2006, \mnras, 373, 705

\bibitem[Ruiz et al.(1997)]{1997ApJ...491L.107R} Ruiz, M.~T., Leggett, S.~K., \& Allard, F.\ 1997, \apjl, 491, L107 

\bibitem[{{Ruiz} {et~al.}(1991){Ruiz}, {Takamiya}, \& {Roth}}]{Ruiz91}
{Ruiz}, M.~T., {Takamiya}, M.~Y., \& {Roth}, M. 1991, \apjl, 367, L59

\bibitem[Salim et al.(2003)]{2003ApJ...586L.149S} Salim, S., L{\'e}pine, S., Rich, R.~M., \& Shara, M.~M.\ 2003, \apjl, 586, L149 

\bibitem[Schmidt et al]{S06} Schmidt, S.J., Cruz, K.L., Bongiorno, B.J., Liebert, J., Reid, I.N., 2007, \aj, 133, 2258

\bibitem[{{Schneider} {et~al.}(2002){Schneider}, {Knapp}, {Hawley}, {Covey},
  {Fan}, {Ramsey}, {Richards}, {Strauss}, {Gunn}, {Hill}, {MacQueen}, {Adams},
  {Hill}, {Ivezi{\'c}}, {Lupton}, {Pier}, {Saxe}, {Shetrone}, {Tufts}, {Wolf},
  {Brinkmann}, {Csabai}, {Hennessy}, \& {York}}]{Schneider02}
Schneider, D.~P. et al., 2002, \aj, 123, 458

\bibitem[Scholz et al.(2002)]{2002MNRAS.329..109S} Scholz, R.-D., Ibata, R., Irwin, M., Lehmann, I., Salvato, M., \& Schweitzer, A.\ 2002, \mnras, 329, 109 

\bibitem[{{Seifahrt} {et~al.}(2005){Seifahrt}, {Guenther}, \&
  {Neuh{\"a}user}}]{Seifahrt05}
{Seifahrt}, A., {Guenther}, E., \& {Neuh{\"a}user}, R. 2005, \aap, 440, 967

\bibitem[Skrutskie et al.(2006)]{2006AJ....131.1163S} Skrutskie, M.~F., et al.\ 2006, \aj, 131, 1163 

\bibitem[{{Tinney}(1996)}]{Tinney96} {Tinney}, C.~G. 1996, \mnras, 281, 644

\bibitem[{{Tinney} {et~al.}(2003){Tinney}, {Burgasser}, \&
  {Kirkpatrick}}]{Tinney03}
{Tinney}, C.~G., {Burgasser}, A.~J., \& {Kirkpatrick}, J.~D. 2003, \aj, 126,
  975

\bibitem[{{Tinney} {et~al.}(1998){Tinney}, {Delfosse}, {Forveille}, \&
  {Allard}}]{Tinney98} {Tinney}, C.~G., {Delfosse}, X., {Forveille}, T., \& {Allard}, F. 1998, \aap,
  338, 1066

\bibitem[{{Tinney} {et~al.}(1993){Tinney}, {Mould}, \& {Reid}}]{Tinney93_TVLM}
{Tinney}, C.~G., {Mould}, J.~R., \& {Reid}, I.~N. 1993, \aj, 105, 1045

\bibitem[{{Tinney} {et~al.}(1995){Tinney}, {Reid}, {Gizis}, \&
  {Mould}}]{Tinney95}
{Tinney}, C.~G., {Reid}, I.~N., {Gizis}, J., \& {Mould}, J.~R. 1995, \aj, 110,
  3014

\bibitem[{{van Altena} {et~al.}(1995){van Altena}, {Lee}, \&
  {Hoffleit}}]{vanAltena}
{van Altena}, W.~F., {Lee}, J.~T., \& {Hoffleit}, E.~D. 1995, {The general
  catalogue of trigonometric [stellar] paralaxes} (New Haven, CT: Yale
  University Observatory, |c1995, 4th ed., completely revised and enlarged)

\bibitem[{{Vrba} {et~al.}(2004){Vrba}, {Henden}, {Luginbuhl}, {Guetter},
  {Munn}, {Canzian}, {Burgasser}, {Kirkpatrick}, {Fan}, {Geballe},
  {Golimowski}, {Knapp}, {Leggett}, {Schneider}, \& {Brinkmann}}]{Vrba04}
Vrba, F.~J. et al., 2004, \aj, 127, 2948

\bibitem[{{Wilson}(2002)}]{Wilson01} {Wilson}, J.~C. 2002, Ph.D.~Thesis, Cornell Univ.

\bibitem[{{Wilson} {et~al.}(2003){Wilson}, {Miller}, {Gizis}, {Skrutskie},
  {Houck}, {Kirkpatrick}, {Burgasser}, \& {Monet}}]{Wilson03}
{Wilson}, J.~C., {Miller}, N.~A., {Gizis}, J.~E., {Skrutskie}, M.~F., {Houck},
  J.~R., {Kirkpatrick}, J.~D., {Burgasser}, A.~J., \& {Monet}, D.~G. 2003, in
  IAU Symposium, Vol. 211, Brown Dwarfs, ed. E.~{Mart{\'{\i}}n}, 197

\bibitem[York et al.(2000)]{2000AJ....120.1579Y} York, D.~G., et al.\ 2000, \aj, 120, 1579 

\end{thebibliography}
\end{document}